\documentclass[
    reprint,
    amsmath,
    amssymb,
    aps,
    prd,
    nofootinbib,
    superscriptaddress,
]{revtex4-2}
    
\usepackage[english]{babel}
\usepackage[utf8]{inputenc}

\usepackage[colorlinks=true,allcolors=blue]{hyperref}
\usepackage[inline]{enumitem}
\usepackage{mathtools}
\usepackage[power-half-as-sqrt]{siunitx}
\usepackage[normalem]{ulem}
\usepackage{orcidlink}
\usepackage{booktabs,dcolumn}

\usepackage{graphicx}
\graphicspath{ {./figures/} }

\newcommand{\tran}[1]{\ensuremath{#1^{\top}}} 

\DeclareMathOperator{\diag}{diag}
\DeclareMathOperator{\Var}{Var}
\DeclareMathOperator{\Cov}{Cov}

\providecommand{\expect}[1]{\ensuremath{\left\langle#1\right\rangle}}

\begin{document}


\title{Recovering unbiased CMB polarization maps using modern ground-based experiments with minimal assumptions about atmospheric emission}

\author{Simon Biquard\,\orcidlink{0000-0002-1493-2963}}
\email{simon.biquard@manchester.ac.uk}
\affiliation{Université Paris Cité, CNRS, Astroparticule et Cosmologie, F-75013 Paris, France}
\affiliation{CNRS-UCB International Research Laboratory, Centre Pierre Binétruy, IRL2007, CPB-IN2P3, Berkeley, California, USA}
\affiliation{Jodrell Bank Centre for Astrophysics, University of Manchester, Oxford Road, Manchester, England M13 9PL, United Kingdom}

\author{Josquin Errard\,\orcidlink{0000-0002-1419-0031}}
\affiliation{Université Paris Cité, CNRS, Astroparticule et Cosmologie, F-75013 Paris, France}

\author{Radek Stompor\,\orcidlink{0000-0002-9777-3813}}
\affiliation{Université Paris Cité, CNRS, Astroparticule et Cosmologie, F-75013 Paris, France}
\affiliation{CNRS-UCB International Research Laboratory, Centre Pierre Binétruy, IRL2007, CPB-IN2P3, Berkeley, California, USA}

\date{\today}

\begin{abstract}
We present a study of unbiased reconstruction of cosmic microwave background (CMB) polarization maps from data collected by modern ground-based observatories.
Atmospheric emission is a major source of correlated noise in such experiments, complicating the recovery of faint cosmological signals.
We consider estimators that require only minimal assumptions about the properties of the unpolarized atmospheric emission, instead exploiting hardware solutions commonly implemented in modern instruments, such as pairs of orthogonal antennas in each focal plane pixel, interleaving of the antenna orientation from pixel to pixel, and polarization signal modulation via a continuously rotating half-wave plate (HWP).
We focus on two techniques: (i) statistical downweighting of low-frequency atmospheric signals and (ii) pair differencing (PD), which involves differencing the signals collected by two detectors in the same focal plane pixel.
We compare their performance and contrast them with the idealized case where the atmospheric signal is perfectly known and can be cleanly subtracted.
We show that PD can be derived from the maximum likelihood principle under very general assumptions about the atmospheric signal, and is therefore optimal in terms of map sensitivity for essentially any such signal.
Remarkably, in the absence of instrumental systematics, but with reasonable variations of detector noise properties, PD yields polarized sky maps with noise levels only slightly worse (typically a few percent) than the ideal reference case.
While the downweighting method could in principle match or surpass this performance, it requires highly accurate models of the atmospheric signal, which are not readily available.
The performance of PD will be affected by instrumental systematics and those leaking the atmospheric signal to the difference time stream may have particularly adverse impact.
However, we expect that some of these effects, as shown in the case of gain mismatch, are efficiently mitigated by a continuously rotating HWP and, therefore, that in the context of modern CMB experimental setups PD could provide a competitive, numerically robust, and efficient practical solution for CMB polarization mapmaking without the need for atmospheric modeling.

\end{abstract}


\maketitle

\section{\label{sec:intro}Introduction}

The cosmic microwave background (CMB) is a powerful cosmological probe that has shaped our understanding of the Universe.
Measurements of its temperature anisotropies have precisely determined the parameters of the standard cosmological model, $\Lambda$CDM.
However, the next frontier in CMB research lies in the study of its polarization anisotropies, which potentially contain the signature of new physics.
The ``$B$-mode'' (parity-odd) component of the spin-2 polarization field is of particular interest and a key target of current and future surveys, such as the Simons Observatory (SO) \cite{SimonsObservatory:2018koc} and LiteBIRD \cite{LiteBIRD:2022cnt}.
On moderate to small angular scales, it is generated by the gravitational lensing effect of large-scale structure acting on the primordial CMB $E$-mode (parity-even) signal.
This effect has been detected by several experiments with high significance in the last decade \cite{POLARBEAR:2017beh, BICEP2:2018kqh, SPT:2015htm}.
However, a large-scale primordial $B$-mode signal is expected if there are significant primordial gravitational waves in the early Universe \cite{Seljak:1996gy}.
The detection of this signal would provide direct evidence for inflation, a period of accelerated expansion that is believed to occur in the early Universe, and predicts the generation of such waves, thus opening a window on extreme high-energy physics \cite{Brout:1977ix,Guth:1980zm,Starobinsky:1980te}.

Ground-based experiments are leading the way in this quest for primordial $B$ modes.
Current measurements from the BICEP/Keck Collaboration have placed an upper limit on the tensor-to-scalar ratio of $r < 0.036$ at $\qty{95}{\percent}$ confidence \cite{BICEP:2021xfz} (a
reanalysis obtained $r < 0.032$ \cite{Tristram:2021tvh}), highlighting the need for improved sensitivity to detect the elusive inflationary $B$-mode signal.
However, achieving constraining measurements of large angular scales from the ground requires overcoming key challenges, including atmospheric contamination and ground pickup.
Atmospheric fluctuations introduce correlated noise that obscures the cosmological signal.
Ground pickup, caused by side-lobe response to thermal radiation from the Earth, further complicates observations by adding an azimuth-dependent contribution that is partially degenerate with the sky signal.
Hardware developments have been implemented to address these challenges, which typically manifest through unpolarized correlated noise.
Successful technologies include dual-polarization detectors and polarization modulators such as rotating half-wave plates (HWPs).

Because the $B$-mode spectrum is much smaller than both the temperature and $E$-mode polarization signals, modern experiments such as the SO deploy tens of thousands of detectors to achieve the required sensitivity.
In this context, we must ensure that the data reduction and analysis pipelines are able to cope with the data volumes generated by these large arrays of detectors, while accurately modeling noise and systematic effects in order to extract the cosmological information distributed over the entire dataset.
\emph{Mapmaking}, which refers to the reconstruction of the sky signal from the time-ordered data (TOD), is a crucial step of this program and one that represents a significant computational burden, since it deals with the full size of the TOD.

This challenge is further amplified by the presence of spurious signals like atmospheric emission or ground pickup.
Consequently, explicit mitigation techniques, such as filtering and/or template deprojection performed at the time stream level, are necessary in order to suppress undesirable contributions.
Such operations typically affect the cosmologically relevant signal in the data, an effect that has to be corrected for.
Different approaches perform this correction at different stages.
For instance, popular ``filter-and-bin'' techniques \cite{BICEP2:2014dgt,POLARBEAR:2014hgp,POLARBEAR:2017beh,SPT-3G:2024atg,CLASS:2023vml,CLASS:2025khf} do so on the power spectrum level.
These have gained widespread popularity due to their flexibility and numerical efficiency, and the ability to deliver unbiased angular power spectra under rather general assumptions for currently achievable survey depths \cite{Hervias2025}.

Other classes of methods attempt to apply the correction directly on the mapmaking stage, and aim to produce unbiased maps of the sky.
Various techniques of this sort have been proposed and applied in the past, ranging from general maximum likelihood to specialized destriping approaches \cite{Stompor:2002jy,Keihanen:2009tj,Dunner:2012vp,Poletti:2016xhi,ElBouhargani:2021umq,CLASS:2023vml}.
In general, they capitalize on an appropriately designed observing strategy to break potential degeneracies between the sky signal and contaminants \cite{Poletti:2016xhi,CLASS:2023vml}.
This is the case, for instance, for contaminants that happen to be synchronous with some well-defined instrumental parameters, such as telescope orientation or HWP rotation.
Such methods are in general more involved in terms of implementation and more costly to execute than the filter-and-bin technique.
As they are based on different assumptions and could potentially deliver more accurate results, they remain of significant interest---an interest that is bound to grow further once the cosmological signals that have eluded us up to now are detected.
From this perspective, atmospheric contamination appears particularly insidious, as it stems from a dynamic, turbulent medium and thus is not easy to characterize in a succinct and sufficiently accurate manner suitable for this class of mapmaking algorithms.

In this work, we study the fidelity and precision of polarization maps produced by unbiased mapmaking methods from data affected by atmospheric contamination, explicitly taking into account typical hardware features of modern experiments.
In particular, we focus on the so-called \emph{pair differencing} technique, applied broadly in past and present experiments; see, e.g., Refs.~\cite{QUaD:2008xpf,BICEP2:2014dgt,POLARBEAR:2014hgp,SPT-3G:2025vtb}.
We show that this method can be derived from maximum likelihood considerations, requiring only a minimal set of assumptions about the atmosphere, and discuss in detail its impact on the sensitivity of the recovered polarization maps obtained from modern CMB experiments.
This work thus complements previous studies that focused predominantly on the differential systematic effects that can affect it \cite{Hu:2002vu, ODea:2006tvb, Shimon:2007au, BICEP2:2015zed, Thomas:2019hak}.

The paper is organized as follows.
In Sec.~\ref{sec:background}, we provide a short introduction to the mapmaking formalism, and discuss specific challenges and features inherent to ground-based experiments.
In Sec.~\ref{sec:polarized-mapmaking} we compare different methods of reconstructing the polarized part of the signal, including the pair differencing technique.
In Sec.~\ref{sec:sensitivity}, we assess the sensitivity of the latter to the predicted inflationary primordial $B$-mode signal.
In Sec.~\ref{sec:systematics}, we discuss differential systematic effects.
Finally, we summarize the results and conclude in Sec.~\ref{sec:conclusion}.

\section{\label{sec:background}Formalism and background}

\subsection{Mapmaking}

\subsubsection{Data model}

We model the TOD $d$ as a \emph{linear} function of the sky signal amplitudes $s$, and additional contributions characterized by amplitudes $a$,
\begin{equation}
    d = P s + T a + n .
    \label{eq:data-model}
\end{equation}    
The \emph{pointing matrix} $P$ and \emph{template matrix} $T$ encode the response of the instrument to $s$ and $a$ respectively.
The time-domain vector $n$ represents instrumental noise, whose average over the statistical ensemble of all possible realizations is assumed to be zero.
The $T a$ term typically describes spurious signals, such as atmospheric emission, which may be nonstationary or strongly non-Gaussian and therefore cannot be captured accurately by the noise model.
Other examples include ``scan-synchronous'' (fixed in horizon coordinates) and ``HWP-synchronous'' signals, which can typically be modeled by templates depending on a well-defined instrumental parameter such as telescope azimuth or HWP angle \cite{Stompor:2002jy, Poletti:2016xhi, ElBouhargani:2021umq}.

A key assumption in Eq.~\eqref{eq:data-model} is that relevant contributions have a limited number of degrees of freedom, i.e., can be described by a limited number of known templates (columns of $P$ and $T$) and corresponding amplitudes (vectors $s$ and $a$).
This implies in particular that we can \emph{discretize} continuous signals (e.g., sky signal) with sufficient accuracy.
For simplicity, we take the same pixelization for all three relevant Stokes parameters $I$, $Q$ and $U$.
Similarly, beams are assumed axially symmetric and the same for all three Stokes components; therefore, they never appear explicitly here and can be thought of as being convolved with the maps.

We also recall that for a perfect linear polarizer coupled to a total power detector, the sky signal part of Eq.~\eqref{eq:data-model} takes the explicit form \cite{WMAP:2003gdj}
\begin{equation}
    (P s)_t = I_{p_t} + \cos(2\varphi_t) Q_{p_t} + \sin(2\varphi_t) U_{p_t} ,
    \label{eq:bolometer-equation}
\end{equation}
where $p_t$ is the observed sky pixel (assuming no pointing interpolation) and $\varphi_t$ is the angle of the polarizer with respect to the sky coordinates, both at time $t$.
Similarly, if an ideal HWP oriented with time-dependent angle $\psi_t$ with respect to the instrument is used to modulate the incoming polarization, the model instead reads \cite{Johnson:2006jk}
\begin{equation}
    (P s)_t = I_{p_t} + \cos(2\varphi_t + 4\psi_t) Q_{p_t} + \sin(2\varphi_t + 4\psi_t) U_{p_t} .
    \label{eq:bolometer-hwp-equation}
\end{equation}

\subsubsection{\label{sec:ml-solution}Maximum likelihood solution}

Let us derive the general \emph{maximum likelihood} (ML) estimator $\hat{s}$ of the sky signal, assuming Gaussian noise with covariance $N$ and a Gaussian prior on $a$.
We may write the corresponding negative log likelihood as
\begin{multline}
    \chi^2 = \tran{(d-Ps-Ta)} N^{-1} (d-Ps-Ta) \\
    + \tran{(a-\bar{a})} \Sigma_a^{-1} (a-\bar{a}) ,
    \label{eq:chi-squared}
\end{multline}    
where $\bar a$ is the expectation value of $a$ and $ \Sigma_a$ is the associated covariance.
The ML solution is found by minimizing $\chi^2$ with respect to $a$ and $s$.
For this, it is useful to change variables to
\begin{equation}
    a' \equiv a - \bar{a}, \qquad d' \equiv d - T \bar{a}
\end{equation}    
such that minimizing the $\chi^2$ with respect to $a$ or $a'$ is equivalent and gives
\begin{equation}
    \hat a' = (\Sigma_a^{-1} + \tran{T} N^{-1} T)^{-1} \tran{T} N^{-1} (d' - Ps) .
\end{equation}    
Injecting this in \eqref{eq:chi-squared} gives
\begin{align}
    \chi^2 &= \tran{(d'-Ps)} \tran{Z} N^{-1} Z (d'-Ps) + \text{const}(s) \\
    \intertext{with}
    Z &\equiv \mathbb{I} - T \left(\Sigma_a^{-1} + \tran{T} N^{-1} T \right)^{-1} \tran{T} N^{-1} .
    \label{eq:projector-with-prior}
\end{align}    
Noticing that $\tran{Z} N^{-1} = N^{-1} Z$, we have the simplification
\begin{equation}
    \tran{Z} N^{-1} Z = N^{-1} Z^2 = N^{-1} Z
\end{equation}    
since $Z$ is a projector.
Finally, after minimizing the $\chi^2$ with respect to $s$, we find
\begin{equation}
    \hat s = (\tran{P} N^{-1} Z P)^{-1} \tran{P} N^{-1} Z (d - T \bar{a}) .
    \label{eq:general-ml-estimate}
\end{equation}

If the prior is improper, i.e., $\Sigma_a^{-1} = 0$, at least for some amplitudes, the corresponding modes are directly deprojected from the original data and not included in the sky signal estimate.
In this extreme case of no prior information, $N^{-1} Z$ becomes the ``filtering and weighting operator'' $F_T$ (cf. Refs.~\cite{Poletti:2016xhi,ElBouhargani:2021umq}),
\begin{equation}
    N^{-1} Z \rightarrow F_T \equiv N^{-1} \left[\mathbb{I} - T (\tran{T} N^{-1} T)^{-1} \tran{T} N^{-1} \right] ,
    \label{eq:filtering-weighting-operator}
\end{equation}
which filters all contributions spanned by the columns of $T$ (i.e., $F_T T = 0$) and weights orthogonal modes by the inverse noise variance.
This construction yields an unbiased estimate of the sky signal in the presence of any systematic effect if $T$ defines a subspace rich enough to encompass every possible manifestation of the effect.
If that subspace is not orthogonal to the sky, i.e., if there are \emph{degeneracies} between $T$ and $P$, the system matrix $(\tran{P} F_T P)^{-1}$ becomes singular, and the corresponding sky modes are impossible to distinguish from the systematic effect and cannot be recovered \cite{Poletti:2016xhi}.
Other modes are by construction unbiased over the statistical ensemble of instrumental noise realizations.
The challenge in this case is to define a sufficiently general $T$ that correctly captures the systematic while keeping the number of potentially degenerate sky modes to a minimum.

When proper priors are available, i.e., when $\Sigma_a^{-1}$ is invertible, one can rewrite $N^{-1} Z$ using the Woodbury matrix identity:
\begin{align}
    N^{-1} Z
    &= N^{-1} - N^{-1} T (\Sigma_a^{-1} + \tran{T} N^{-1} T)^{-1} \tran{T} N^{-1} \nonumber \\
    &= (N + T \Sigma_a \tran{T})^{-1} \equiv \widetilde N^{-1} .
    \label{eq:combined-noise-covariance}
\end{align}    
The ML estimate now reads
\begin{equation}
    \hat s = (\tran P \widetilde N^{-1} P)^{-1} \tran P \widetilde N^{-1} (d - T \bar{a}) ,
    \label{eq:wghtEstimGen}
\end{equation}    
where $\widetilde N$ includes both the variance due to instrumental noise and the systematic contribution, which is now characterized statistically around its average.
We refer to this as the \emph{downweighting} (DW) approach hereafter.
This estimator is unbiased over the ensemble of noise and systematic effect realizations, if only the average signal, $\bar{a}$, is known.
In many actual applications, it is assumed to vanish.
The challenge then is to find a representation of the total covariance $\widetilde N$ that is statistically sufficient and computationally manageable.

As emphasized earlier, our sole focus will be the recovery of the polarization sky signal, i.e., maps of the $Q$ and $U$ Stokes parameters, from the available data.
Typical datasets are composed of measurements taken by total power detectors, hence the measured signals combine all three Stokes parameters, $I$, $Q$, and $U$ (neglecting circular polarization).
The relevant pointing matrix  can be split into polarized and total intensity parts, as well as the corresponding sky map:
\begin{equation}
    P = \begin{bmatrix}
        P_I & P_{QU}
    \end{bmatrix} ; \quad s = \begin{bmatrix}
        s_I \\ s_{QU}
    \end{bmatrix} .
\end{equation}
While we can in principle estimate the full sky map $s$ (and possibly drop its $I$ component \emph{a posteriori}), a somewhat more general approach is to marginalize over the \emph{a priori} unknown total intensity part by deriving the estimators directly for $Q/U$ only.
This is also applicable to situations where the total intensity signal is not recoverable.
The relevant expressions can be readily obtained in the formalism presented above by extending the set of templates $T$,
\begin{equation}
T \longrightarrow T' = \begin{bmatrix} T & P_I \end{bmatrix},
\end{equation}
and assuming improper priors for all the extra degrees of freedom, such that
\begin{equation}
    \hat s_{QU} = \left(\tran{P}_{QU} N^{-1} Z' P_{QU} \right)^{-1} \tran{P}_{QU} N^{-1} Z' d,
    \label{eq:ponly-ml-estimate}
\end{equation}
where $Z'$ follows Eq.~\eqref{eq:projector-with-prior} with the extended templates $T'$.

Lastly, we note that different choices of templates can be used to model a given systematic effect and that this choice may depend on the mapmaking method as well as the specific model of the data and the details of the experiment and its operations. We discuss this in the next section.

\subsection{Ground-based CMB observations}

\subsubsection{\label{sec:atmosphere}Atmospheric emission}

Ground-based telescopes have to observe the sky through the Earth's atmosphere.
While the latter has regions of reduced opacity in the $30-\qty{300}{\GHz}$ range where the CMB is at its brightest, even in these atmospheric windows there remains significant emission.
This radiation increases the optical loading of the detectors and therefore raises their photon noise level, limiting the overall sensitivity of the instrument.
More critically, fluctuations in the water vapor density column cause spatial and temporal variations in the atmosphere's brightness temperature, which introduce correlated noise both in time and between detectors.
Those fluctuations are typically the dominant source of low-frequency noise for ground-based experiments \cite{Lay:1999qi,ACBAR:2002eoj,POLARBEAR:2015mbo,ACT:2025xdm} and represent one of the primary challenges for sensitive large-scale CMB measurements from the ground.

While atmospheric emission is slightly circularly polarized due to Zeeman splitting of oxygen in the Earth's magnetic field, the expected linear polarized intensity is very low \cite{Hanany:2003ms,Spinelli:2011fb}.
Previous studies have set a $\qty{1}{\percent}$ limit on the linear polarization fraction of atmospheric emission \cite{POLARBEAR:2015mbo}.
However, the presence of horizontally aligned ice crystals in tropospheric clouds can generate spurious polarized emission by scattering radiation from the ground \cite{POLARBEAR:2018xcv,CLASS:2023mlm,SPT-3G:2024yab}.
This causes significant “bursts” of horizontal (Stokes $Q < 0$) polarization in the data, which manifest as excess low-frequency noise in the TOD and are not mitigated by polarization modulation.
Possible avenues for addressing this issue have been discussed in, e.g., Ref.~\cite{SPT-3G:2024yab}.
We will ignore this effect in the present work.
Therefore, for the rest of the paper, we model the atmosphere as an \emph{unpolarized} signal correlated both spatially and in time.
This opens multiple avenues to mitigate its impact on the recovered $Q$ and $U$ maps through combinations of appropriate hardware and software solutions.

\subsubsection{\label{sec:hardware-solutions}Hardware solutions}

Ground-based CMB experiments have developed specialized hardware components to allow for an efficient detection of polarized sky signals in the presence of correlated noise sources.
These help to address the problem of atmospheric contamination.
Two key technologies have proved particularly useful: dual-polarization detectors  and polarization modulators (notably HWPs).

Dual-polarization detectors allow us to measure both the total power $I$ and a linear combination of $Q$ and $U$ in a single focal plane pixel thanks to two orthogonal antennas \cite{Montroy:2005yx, Yoon:2006jc, Jones:2006ac, Kuo:2008juj, QUaD:2008una, BICEP3:2014snp, Henderson:2015nzj, Posada:2015yey, Stebor:2016hgt, SimonsObservatory:2018koc}.
Samples collected by orthogonal antennas can be \emph{differenced} in order to reject unpolarized signals, whose intensity is independent of antenna orientation.
This enables the reconstruction of polarized signals (antenna angle dependent) with minimal contamination if the antennas are close to orthogonal, and also with negligible or small loss of precision (cf. Sec.~\ref{sec:sensitivity}).

Polarization modulators perform \emph{frequency modulation} to shift the polarization signal to higher frequencies in the TOD.
With sufficiently high modulation frequency, the signal is measured in a band where detector sensitivity is mostly limited by photon noise rather than atmospheric fluctuations.
The most common implementation of this technique is a continuously rotating HWP, which modulates the polarization signal at four times its mechanical rotation frequency.
HWPs have been deployed and characterized for several millimeter-wave polarization experiments \cite{Johnson:2006jk,ABS:2013dqh,Ritacco:2016due,Takakura:2017ddx}. Despite obvious advantages, they can also introduce their own systematic effects \cite{Johnson:2006jk,ABS:2013dqh,ABS:2016rpo}, including intensity-to-polarization
leakage, modulation efficiency variations across the bandwidth, and mechanical vibrations.
These effects require careful calibration and modeling in the data analysis pipeline but are out of the scope of this paper.

\section{\label{sec:polarized-mapmaking}Polarized mapmaking for ground-based experiments}

We focus hereafter on unpolarized atmospheric emission as the only relevant systematic effect and aim exclusively to recover the polarized sky signals from the data.

For this purpose, we study and compare three approaches corresponding to varying degrees of knowledge and assumptions about the atmospheric signal:
\begin{enumerate}
    \item A \emph{minimal} model where we only assume that the atmospheric contamination is the same for both detectors of any orthogonal pair in the focal plane (cf. Sec.~\ref{sec:minimal-model}).
    \item A \emph{statistical} model based on the downweighting approach where we assume that the combined noise covariance \eqref{eq:combined-noise-covariance}, accounting for both instrumental noise and atmospheric fluctuations, can be described as stationary over defined time intervals, with the noise power spectral density (PSD) derived from the TOD itself (cf. Sec.~\ref{sec:statistical-model}).
    \item An \emph{idealized} case where we assume that the atmospheric contribution is perfectly known.
    In the formalism of Sec.~\ref{sec:ml-solution}, this corresponds to $\Sigma_a \rightarrow 0$ and $a \rightarrow \bar{a}$, and the ML solution is simply the standard estimate in the absence of any atmospheric contamination, since it can be readily subtracted:
    \begin{equation}
        \hat s^\text{ideal} \equiv (\tran{P} N^{-1} P)^{-1} \tran{P} N^{-1} d^\text{atm-free} .
        \label{eq:ideal-estimate}
    \end{equation}
    While this case is unattainable in practice, it does provide a valuable statistical benchmark.
\end{enumerate}

Before going any further, we note that, in the presence of a rotating HWP, specialized mapmaking approaches based on explicit, ``lock-in'' \emph{demodulation} of the data can recover $Q$ and $U$ Stokes parameters efficiently (see, e.g., Refs.~\cite{Johnson:2006jk,ABS:2013dqh,Ritacco:2016due} for more details) by isolating the polarization information from low-frequency noise in the time streams.
In principle, this method also has the advantage that polarization can be reconstructed from individual detectors in isolation (not relying on pairs of orthogonal antennas), which can potentially help when dealing with differential systematics that affect the two antennas of a pair differently, such as beam mismatch or pointing errors.
However, the combination of the information from multiple detectors is necessary, albeit at a later stage of the analysis, to produce statistically optimal results.
Discussions about demodulation and its interaction with multiple detectors can be found in Refs.~\cite{Brown:2008ha,Rashid:2023xnw}.

In this work, we consider an alternative approach and do not assume any data preprocessing.
Instead, an effective demodulation is only performed as an integral part of the mapmaking procedure, as encoded in the pointing matrix, Eqs.~\eqref{eq:bolometer-equation} and \eqref{eq:bolometer-hwp-equation}, and thus its impact on the sky signal is correctly included and corrected for.
This also allows us to use the same formalism and implementation for both HWP and non-HWP cases and to compare the two approaches more straightforwardly.

\subsection{\label{sec:minimal-model}Minimal model}

A fundamental challenge in atmospheric mitigation is the unpredictable and complex nature of its emission.
If the atmosphere were fixed in Earth-bound coordinates, modeling its fluctuations over moderate timescales on a two-dimensional map of pixels fixed to this coordinate system (different from sky coordinates) could be a promising avenue toward its characterization and removal, given that it typically only varies on rather large angular scales.
In practice, however, the presence of wind means that the atmosphere is potentially only ``fixed'' in coordinates moving with the wind.
As the wind's speed is \emph{a priori} unknown, the effect of the atmosphere can not be easily captured by a linear model such as Eq.~\eqref{eq:data-model}.

Instead, to work around this problem, we may want to consider a construction allowing for maximal flexibility in the atmospheric modeling.
An extreme example would be a model with as many degrees of freedom as measured data points, i.e., $T = \mathbb I$ (the identity matrix) with dimensions given by the total number of samples.
This obviously would exceed the number of available constraints, making the problem degenerate.
However, we can break this degeneracy if the telescope is equipped with dual-polarization detectors, making the only assumption that unpolarized signals are measured identically by orthogonal antennas in a pair and assuming no priors ($\Sigma_a^{-1} = 0$) on the corresponding amplitudes, thus making a minimal number of assumptions on the atmospheric signal.
Every column of $T$ now has two nonzero entries, corresponding to the two detectors of a pair, such that $T$ is conceptually two identity matrices stacked on top of each other for every focal plane pixel:
\begin{equation}
    T = \begin{bmatrix} \mathbb I \\ \mathbb I \end{bmatrix} .
    \label{eq:minimal-template}
\end{equation}
As discussed in Sec.~\ref{sec:ml-solution}, this yields a sky estimate that is free of signals captured by $T$.
Note that this choice of $T$ captures sky total intensity as well since $T P_I = P_I$, so only the polarized sky components are estimated.

In Appendix~\ref{sec:append-marginalization}, we show that this minimal model leads in fact to the well-known \emph{pair differencing} estimator, which we now define.
For any pair of orthogonal detectors denoted by $\parallel$ and $\perp$ subscripts, we are free to transform the TOD into \emph{sum} and \emph{difference} data,
\begin{equation}
    \begin{bmatrix} d_+ \\ d_- \end{bmatrix}
    \equiv \frac12
    \begin{bmatrix}
        d_\parallel + d_\perp\\
        d_\parallel - d_\perp
    \end{bmatrix} .
\end{equation}
We emphasize that this transformation does not by itself entail any loss of information, as it is represented by the block matrix
\begin{equation}
    X \equiv \frac12 \begin{bmatrix*}[r]
        \mathbb I & \mathbb I \\
        \mathbb I & -\mathbb I
    \end{bmatrix*}
    \quad \text{with inverse} \quad
    X^{-1} = 2 X .
    \label{eq:transformation-matrix}
\end{equation}
Any estimate previously expressed in terms of $d$ could be equivalently expressed in terms of the transformed dataset, $Xd$.

However, since the difference data, $d_-$, is free of atmospheric signal (ideally) and contains all the polarization information, it seems natural to consider it an independent data vector and process it by itself to recover the polarization signal.
We thus call pair differencing (PD) estimate the following $Q/U$ estimator, computed from $d_-$, and discarding $d_+$,
\begin{align}
    \hat s^\text{pd} \equiv \hat s^\text{pd}_{QU}
    &\equiv (\tran{P_-} N_-^{-1} P_-)^{-1} \tran{P_-} N_-^{-1} d_- ,
    \label{eq:pd-estimate}
\intertext{where}
    N_- &\equiv \expect{n_- \tran{n_-}}
\intertext{is the covariance of the difference noise, and}
    P_- &\equiv \frac12 (P_\parallel^{QU} - P_\perp^{QU})
\end{align}
is the relevant part of the pointing matrix for the difference time stream.

As part of the available information is discarded, it would seem that this approach is suboptimal.
However, we showed above that it results from a rigorous maximum-likelihood derivation and is therefore \emph{as optimal as only possible} given the assumptions made about the atmospheric signal.
We discuss this further in Sec.~\ref{sec:sensitivity}.

In practice, we estimate $N_-$ in Eq.~\eqref{eq:pd-estimate} from the data, and this may differ from the true covariance.
Our implementation of the PD estimator is then given by
\begin{equation}
    \hat s^\mathrm{pd}_W \equiv (\tran{P_-} W P_-)^{-1} \tran{P_-} W d_- .
    \label{eq:pd-estimate-gen}
\end{equation}
This expression is manifestly unbiased but may not always reach the same precision as Eq.~\eqref{eq:pd-estimate}.
The general expression for the covariance of the estimated maps is
\begin{equation}
    \hskip -0.2truecm \mathcal{N}_{QU}^\mathrm{pd} \equiv (\tran{P_-} W P_-)^{-1} \;
    \tran{P_-} W N_- W P_- \;
    (\tran{P_-} W P_-)^{-1} .
    \label{eq:pd-covariance-gen}
\end{equation}

\subsection{\label{sec:statistical-model}Statistical downweighting}

For comparison, we also consider an alternative approach motivated by Eq.~\eqref{eq:wghtEstimGen}, and treat the atmospheric signal as a stochastic contribution, assuming that on average it vanishes, i.e., $\bar a = 0$.
In doing so, we accept the fact that the atmosphere will contribute to our final sky signal estimates; however, we hope to minimize its impact by appropriately downweighting the modes where its contribution is significant, and then to be able to account for the extra uncertainty in the final map error budget.
Reaching both of these goals requires finding a suitable description of the atmospheric signal covariance.

It has been shown that distinguishing atmospheric modes from systematics in the data involves somewhat sophisticated TOD processing \cite{Morris:2024etq}.
Thus, successful models may require advanced constructions (see, e.g., Ref.~\cite{Dunner:2012vp}), in particular if one aims to recover the total intensity of the CMB.
Consequently, such models could be numerically complex and not straightforward to validate.
As our focus is solely on polarization, we expect, however, that simpler and less demanding models could be already sufficient.
In particular, we will neglect noise correlations that are induced by the atmosphere between different detectors \cite{POLARBEAR:2015mbo}.
Our DW estimator then reads
\begin{equation}
    \hat s^\text{dw} 
    \equiv (\tran{P} W P)^{-1} \tran{P} W d .
    \label{eq:dw-estimate-gen}
\end{equation}
Contrary to the PD estimator \eqref{eq:pd-estimate-gen}, it uses the full dataset $d$ as an input and includes all three Stokes parameters in the pointing matrices and the recovered map.
While we could marginalize right away over the total intensity part, in practice, the resulting equations are more cumbersome to implement, and there is little advantage in doing so, if any, from the perspective of numerical efficiency.
We thus simply drop the total intensity map at the end of the mapmaking procedure.

We note that the inversion of the system matrix in Eqs.~\eqref{eq:pd-estimate-gen} and \eqref{eq:dw-estimate-gen} may fail if some pixels are not observed with sufficient redundancy.
In particular, to estimate the $Q$ and $U$ amplitudes in a given pixel, it is necessary to observe it with at least two sufficiently different ``crossing angles'', i.e., effective antenna orientations (taking HWP rotation into account).
Whenever these conditions are not met, the corresponding samples need to be flagged and excluded from the reconstruction, along with those contaminated by glitches, or acquired during unstable scanning intervals.
The treatment of such gaps in the time stream requires dedicated procedures \cite{Stompor:2002jy}.

The weights $W$ adopted in Eq.~\eqref{eq:dw-estimate-gen} assume no detector-detector correlations.
They are thus block diagonal, with each detector-specific block given by the PSD of the detector time stream.
While this is probably highly suboptimal for the recovery of the total intensity maps, one may hope that it is sufficient for polarization.
This is one of the questions we investigate in the following.
The error covariance of the estimated maps is then given by
\begin{equation}
    \mathcal{N}^\text{dw} \equiv
    (\tran{P} W P)^{-1} \;
    \tran{P} W \widetilde{N} W P \;
    (\tran{P} W P)^{-1},
    \label{eq:dw-covariance-gen}
\end{equation}
where $\widetilde{N}$ is the covariance of the instrumental noise and the atmospheric signal combined together; cf. Eq.~\eqref{eq:combined-noise-covariance}.
As we target polarization maps here, only the $Q/U$ blocks of $\mathcal{N}^\text{dw}$ are of actual interest.
One can write down an analytic expression for them; however, it is long and rather complex and thus not very enlightening.

Instead, let us try and draw some conclusions directly from Eq.~\eqref{eq:dw-covariance-gen}.
The estimated maps are minimum variance if $W = \widetilde{N}^{-1}$, and any other choice will unavoidably lead to increased uncertainty.
As the atmosphere contribution is unpolarized, it only affects the $I \times \{I,Q,U\}$ blocks of the central product $\tran{P} W \widetilde{N} W P$.
Its impact on the polarization maps can be mitigated if the $I \times Q/U$ blocks of both $\tran{P} W P$ and $\tran{P} W \widetilde{N} W P$ vanish.
This latter requirement ensures that the atmospheric signal is well separated from the polarization signal, and the former ensures that the variance due to atmosphere is never folded back in the polarization maps' covariance.
There is a potential huge benefit to ensuring that those two conditions are met; in the following, we discuss how to achieve this by adapting mapmaking choices to the hardware features discussed earlier.
Nonetheless, we also show that those conditions are not sufficient to ensure the optimality of the polarization estimates, and that weighting may need to be adjusted accordingly.

\subsection{Comparison of both approaches}

We present here a comparison of the discussed map-making approaches, PD and DW, in order to demonstrate salient features of their performance.

\subsubsection{\label{sec:simulations}Simulations}

As we focus on the recovery of large-scale polarization, we consider an SO-like small aperture telescope (SAT) observing at $\qty{90}{GHz}$, equipped with pairs of orthogonal detectors and a rotating HWP.
The time-domain simulations are generated using the \texttt{TOAST 3}\footnote{\url{https://github.com/hpc4cmb/toast/tree/toast3}} framework and the \texttt{sotodlib}\footnote{\url{https://github.com/simonsobs/sotodlib}} library.
The reduction of the data into HEALPix maps \cite{Gorski:2004by} is performed with the \texttt{MAPPRAISER}\footnote{\url{https://github.com/B3Dcmb/midapack}} library \cite{ElBouhargani:2021umq}.
The resolution parameter is $N_\mathrm{side} = 512$.

The simulated instrument scans a region of the sky south of the Galactic plane, similar to the ``SAT South field'' described in Ref.~\cite{Stevens:2018biw}.
Data are acquired with a sample rate of $\qty{37}{Hz}$.
The HWP rotation frequency is set to $\qty{120}{rpm}$ and can be turned off.
The dataset consists of $\sim\qty{1}{\hour}$ long constant-elevation scans (CESs) taken over the course of one day.
To reduce computational requirements, we only keep one in four focal plane pixels, resulting in roughly 700 detectors.
The total size of the dataset (including TOD, pointing information, etc.) is measured to be around $\qty{240}{\giga\byte}$.

As the mapmaking procedure that we use is linear, we directly obtain noise maps by only simulating realistic atmosphere signal and instrumental noise.
There are no systematics (gain errors, beam mismatches, etc.).
The atmosphere simulation in \texttt{TOAST} is based on a physical model of the  atmosphere \cite{POLARBEAR:2015mbo} and uses a three-dimensional structure of Kolmogorov turbulence moving through the telescope’s field of view, with a distribution of temperature, water vapor, and wind speed.
On the other hand, instrumental noise streams (uncorrelated between detectors) are simulated according to a $1/f$ model,
\begin{equation}
    S(f) = \sigma^2 \left[ 1 + \left( \frac{f}{f_\mathrm{knee}} \right)^\alpha \right] ,
    \label{eq:noise-model}
\end{equation}
with nominal parameter values typical of detector noise,
\begin{equation}
    \bar\alpha = -1.0 \quad\text{and}\quad \bar f_\mathrm{knee} = \qty{50}{\milli\Hz} .
    \label{eq:nominal-params}
\end{equation}
The nominal value of the noise equivalent temperature, $\sigma$, does not matter since we will be always be looking at results \emph{relative} to a reference case.

In the nominal case, all detectors get the same instrumental noise parameters.
Alternatively, we added the possibility of varying parameters across the focal plane.
In this case, each parameter is perturbed around its nominal value by a random multiplicative sample $\sim \mathcal N(1, z^2)$.
We will usually quote the dispersion $z$ in percent.
A different set of perturbations is applied for each CES and detector.

\subsubsection{Results}

Using these simulations, we perform mapmaking runs following the three models presented earlier, recovering $Q$ and $U$ noise maps.

\begin{figure}
    \includegraphics[width=\columnwidth]{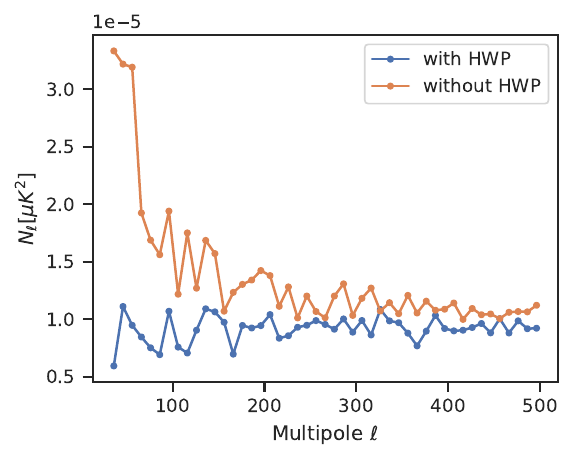}
    \caption{
        Reference $BB$ noise power spectra obtained from the ideal reconstruction (atmosphere-free, only instrumental noise).
    }
    \label{fig:reference-spectra}
\end{figure}

As a figure of merit, we compute for each case the noise power spectra, i.e., the angular power spectra of the noise maps.
We then compare them with \emph{reference} spectra, i.e., the noise spectra of the ideal runs, shown in Fig.~\ref{fig:reference-spectra}.
All spectra are obtained using the \texttt{NaMaster}\footnote{\url{https://github.com/LSSTDESC/NaMaster}} package \cite{Alonso:2018jzx}, with a bin size of 10, and keeping only bins whose center multipole $\ell_b \in [30, 500]$.
In addition, for each pixel $p$, we also compute the diagonal block of the white noise covariance,
\begin{align}
    \Delta &\equiv (\tran P \diag N^{-1} P)^{-1}.
    \label{eq:white-noise-covariance}
\end{align}
These blocks have dimensions $3\times3$ for the downweighting approach and $2\times2$ for pair differencing, corresponding to the number of relevant Stokes parameters.
For each pixel, they are explicitly given by\footnote{
    Note that $P_{t p s} = 0$ whenever $p$ is not the sky pixel observed at time $t$.
}
\begin{align}
    (\Delta_p^{-1})_{s s'} &= \sum_t P_{p t s} N^{-1}_{t t} P_{t p s'},
    \label{eq:cov-block}
\end{align}
where the subscripts $s, s' \in \{I,Q,U\}$ stand for a Stokes parameter.
These blocks quantify the coupling of the noise map between the different estimated Stokes parameters.

\begin{figure*}
    \includegraphics[width=0.8\textwidth]{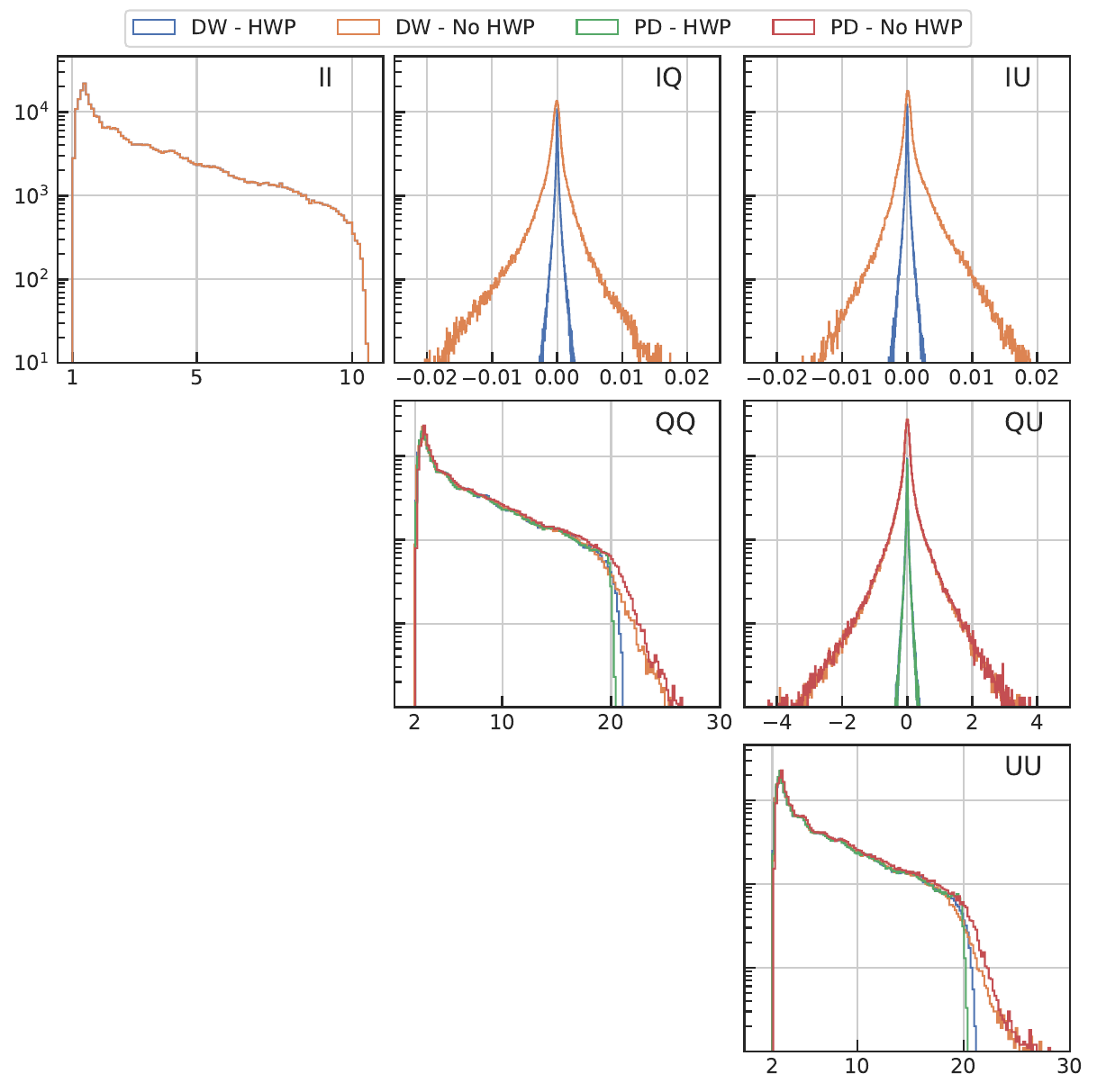}
    \caption{\label{fig:blocks}
        Comparison of white noise covariance block values for DW/PD and HWP/no-HWP cases.
        Nominal values of the instrumental noise parameters are used.
        Noise weights are fitted to the data independently for each detector.
        Each panel of this $3\times3$ histogram matrix shows the distribution of values of the white noise covariance blocks \eqref{eq:cov-block} across the map.
        These blocks are normalized by the variance of the best constrained pixel, such that their smallest values across the map are 1 for $II$ and 2 for $QQ$ and $UU$.
    }
\end{figure*}

For the cases studied here, we display the elements of the blocks in Fig.~\ref{fig:blocks}.
As discussed in Sec.~\ref{sec:statistical-model}, in DW mapmaking, off-diagonal elements $IQ$ and $IU$ are particularly important as they determine the contribution of the atmospheric signal variance to the polarization maps.
When noise weights are fitted independently for each detector, the values of these elements, though centered at zero, scatter significantly around it.
As expected, the scatter is particularly pronounced without HWP; however, in both cases, this is bound to increase the uncertainty of the derived maps, given the magnitude of the atmospheric signal.
While the effect can be further limited by imposing strict selection criteria on retained pixels, this would unavoidably lead to a loss of observed sky fraction, thereby enhancing the sample variance on the largest angular scales.

A more fruitful approach would be to find a way to suppress these off-diagonal elements.
This can be done in a robust way through enforcing the noise weights for two detectors in each pair to be the same. While this will make the weighting of each detector stream less optimal, the overall gain may still be appreciable.
In the following, we confront these conclusions with the results of our simulations.

\begin{figure*}
    \includegraphics[width=\textwidth]{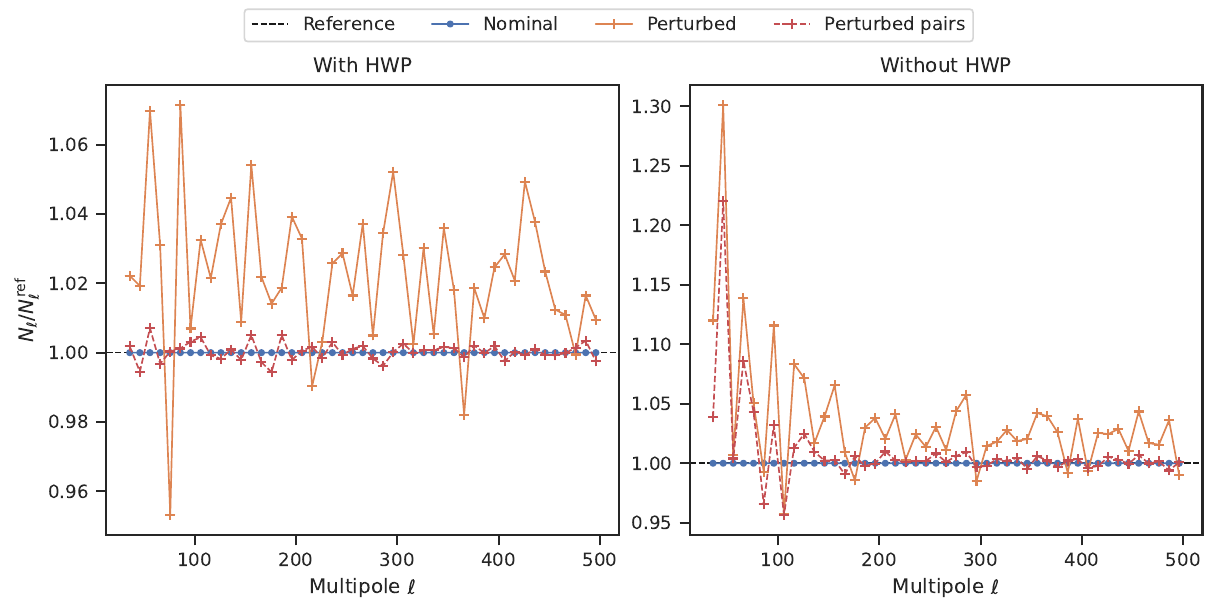}
    \caption{\label{fig:pd-spectra}
        Comparison of $BB$ noise power spectra using pair differencing, with (left) vs. without (right) HWP rotation.
        Different setups are represented with combinations of markers and colors.
        ``Nominal'' (blue) corresponds to the nominal instrumental noise model.
        ``Perturbed'' (orange) has instrumental noise parameters drawn with a dispersion of $\qty{10}{\percent}$ around nominal values.
        ``Perturbed pairs'' (red) is a case where perturbations are applied in such a way that detectors of one pair always have the same parameters, i.e., any variations are only between different detector pairs.
        For each case, we plot the ratio of the noise power spectrum over that of the corresponding ideal (reference) case.
    }
\end{figure*}

\begin{figure*}
    \includegraphics[width=\textwidth]{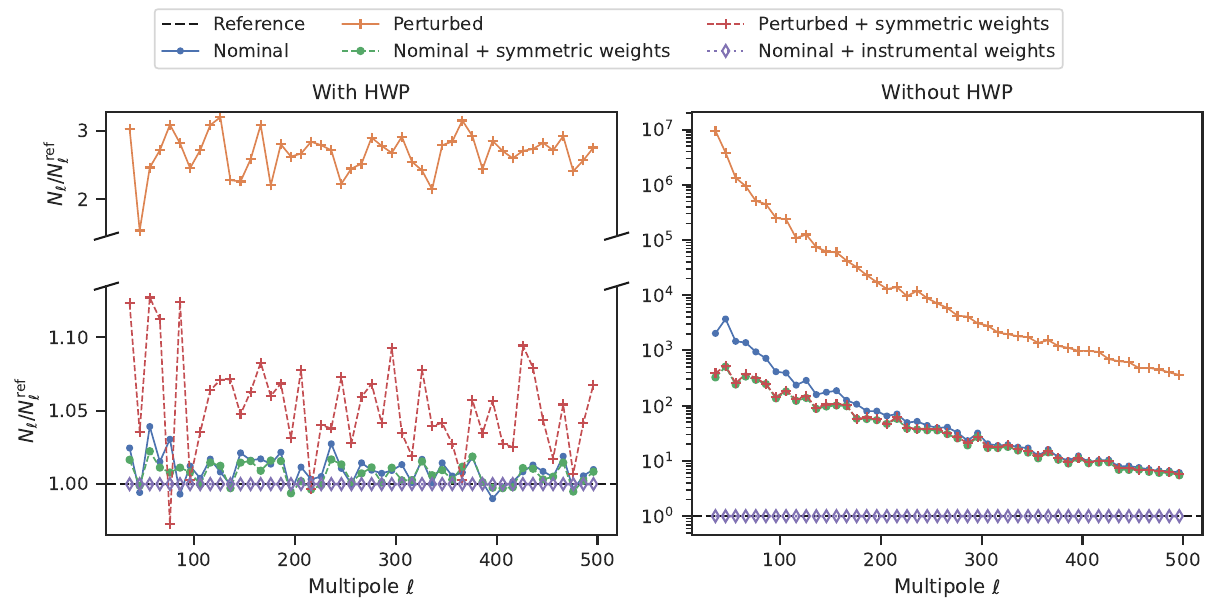}
    \caption{\label{fig:dw-spectra}
        Comparison of $BB$ noise power spectra using downweighting, with (left) vs without (right) HWP rotation.
        Different setups are represented with combinations of markers and colors.
        ``Nominal'' and ``Perturbed'' cases are the same as in Fig.~\ref{fig:pd-spectra}.
        ``Nominal + symmetric weights'' is a case where instrumental noise parameters are nominal, and moreover the assumed noise weights are forced to be symmetric, i.e., the same for both detectors of a pair.
        ``Perturbed + symmetric weights'' is the same idea but with the perturbed noise parameters.
        ``Nominal + instrumental weights'' has both nominal instrumental parameters and noise weights following that model (\emph{not} fitted to the data which contains atmosphere).
        For each case, we plot the ratio of the noise power spectrum over that of the corresponding ideal (reference) case.
    }
\end{figure*}

The resulting $BB$ noise spectra are shown in Figs.~\ref{fig:pd-spectra} and \ref{fig:dw-spectra} ($EE$ is not shown as the results are nearly identical).
The key observations are as follows.
For PD, Fig.~\ref{fig:pd-spectra}, they coincide perfectly with those of the ideal case if the noise properties of both detectors in each pair are the same.
This is independent from the noise differences between the pairs as indeed expected given that these are optimally weighted in the procedure (see Ref.~\cite{ElBouhargani:2021phd} and Sec.~\ref{sec:sensitivity}).
If, on the contrary, the noise properties of detectors in a pair differ, PD yields noisier than ideal maps.
For the $z = \qty{10}{\percent}$ dispersion of parameters in the simulations, the precision loss seems to be $\sim\qty{2}{\percent}$.
We investigate this issue in more depth in Sec.~\ref{sec:sensitivity}.
The main impact of having a HWP is that the noise increase is essentially constant across all multipoles, all the way to the largest angular scales considered.
If no HWP is used, the increase is more pronounced at the low end of the multipole range, all the more so when noise parameters vary, with the additional large-scale variance becoming apparent at $\ell \sim 100 \div 150$.
However, the total increase of power at low multipoles over the white noise plateau is not more than $\sim \qty{30}{\percent}$.
We note that in all these cases we used the same set of map pixels, even if in cases without HWP these are generally less well conditioned than in cases with HWP, as shown in Fig.~\ref{fig:blocks}.

As far as DW mapmaking is concerned, the situation is somewhat more complex.
Let us start by inspecting the cases where HWP rotation is enabled (left panel of Fig.~\ref{fig:dw-spectra}).
Even for the simplest case of nominal noise parameters, the noise spectra do not exactly match those of the ideal case.
The difference is small but persistent, and it does not disappear even when we symmetrize the weights within each detector pair, as shown by the blue (nominal weighting) and green curves (symmetrized weighting), distinct from the black dashed line.
This effect goes away only if we use noise weights with a lower $f_\mathrm{knee}$, e.g., similar to instrumental noise, Eq.~\eqref{eq:nominal-params}, as represented by thin diamond-shaped markers.
On the other hand, if we simulate different noise levels (and use corresponding weights) for each detector, we see significant increase of the noise (orange curve).
We interpret these observations as follows.
When noise weights are (nearly) identical within each pair of detectors, the atmosphere does not contribute directly to the variance of the polarization maps.
Indeed, the mapmaking operator performs \emph{implicit} pair differencing while computing $\tran{P} W d$.
Nevertheless, since the weights account for atmospheric contamination but neglect that it is highly \emph{correlated} between detectors, they result in suboptimal maps.
The effect is small, owing to the HWP modulation, which mitigates this low-frequency noise by shifting the polarization signal well above the atmospheric $f_\mathrm{knee}$.
Still, we find an excess $\sim\qty{1}{\percent}$ variance at the power spectrum level, but this goes away if we decrease sufficiently the assumed $f_\mathrm{knee}$ such that the signal in the polarization band is weighted uniformly.
The method then matches the performance of the reference and PD cases.

These conclusions are fully consistent with the noise spectra obtained in the cases without HWP (right panel of Fig.~\ref{fig:dw-spectra}).
The impact of atmosphere is, however, generally enhanced by the presence of more significant leakage of the atmosphere into polarization, as expected from Fig.~\ref{fig:blocks}.
In the case of detector-dependent noise parameters (orange curve), the mismatch of weights within each detector pair further prevents the implicit differencing mentioned above, leading to huge atmospheric noise residuals dominating on all angular scales by orders of magnitude.
If we now enforce symmetric weights (red curve), the noise spectra are suppressed by 3 to 4 orders of magnitude and become comparable to those from cases with nominal noise parameters (blue and green curves).
Indeed, the corresponding $3\times3$ blocks and the noise weights are now expected to be similar.
The ``symmetrized'' cases show that the remaining extra noise power is purely due to suboptimal weighting, since the atmosphere is removed as a result of implicit differencing.
The noise weights in those cases include the atmospheric $1/f$ and therefore give much less weight to Fourier modes below the atmospheric $f_\mathrm{knee}$, where most of the polarization signal resides in the absence of HWP modulation.
This in turn leads to significantly increased variance at large multipoles.
Like before, we can reduce this uncertainty to the reference level by simply taking noise weights as if they were given by the instrumental noise only (diamond markers).
Alternately, we could improve the noise model by allowing for correlations between detectors in the same focal plane pixel.

We thus conclude that in these conditions the most beneficial setup for downweighting is to assume identical weights within each pair of detectors and derive them assuming the properties of the instrumental noise.
It was indeed shown in real conditions that assigning both detectors of a pair the same low-frequency weights, derived from the pair-difference TOD, leads to substantial improvement of the large-scale polarized noise \cite{SPT-3G:2025vtb}.
But then the best that DW can do is to reproduce the results of pair differencing, the latter being in addition more numerically efficient and robust.
We expect this conclusion to hold for as long as there is no better and more reliable model of the atmospheric signal available, which would allow DW to benefit from optimal propagation of the instrumental noise, as in the ideal case, while keeping the atmospheric leakage under control.
This is, however, a tall order as argued in Sec.~\ref{sec:minimal-model}.
Meanwhile, the PD approach, thanks to its independence from the atmosphere model, may offer a competitive and practical option.
However, as it is, if we are ready to accept a small sensitivity hit, then for the experiments with continuously rotating HWP and pairs of orthogonal detectors, the downweighting approach can still achieve very good performance if only symmetric weights are adopted.
Moreover, if the atmospheric signal is somehow suppressed prior to actual mapmaking, then this symmetrization may not even be needed, allowing the method to benefit from more optimal weighting of individual detectors.
This idea is implemented, for instance, in demodulation methods \cite{Johnson:2006jk,Kusaka:2018yzq,Rashid:2023xnw}.

\section{\label{sec:sensitivity}Sensitivity assessment of the pair differencing approach}

In this section, we address the question of \emph{sensitivity} (or precision) of the pair differencing method.
We argued in Sec.~\ref{sec:minimal-model} that PD is an optimal polarization estimator under the single assumption that unpolarized signals are measured identically by detectors in a pair.
However, we should in general expect it to yield a lower signal-to-noise ratio than the unattainable ideal case \eqref{eq:ideal-estimate} where the atmospheric contribution is perfectly known and subtracted from the data.
Given the very general assumption behind the PD derivation, we might even expect a significant loss of precision: the number of templates that it implicitly marginalizes over, and thus the number of additional degrees of freedom, is as large as half the length of the total combination of data from all the detectors.

However, we already saw in several specific examples of the previous section that the PD estimator is ideal when both detectors in a pair have \emph{identical noise properties}.
In Appendix~\ref{sec:append-iqu-pol-estimator}, we show rigorously why this is the case and obtain the expression for the polarized part of the ideal estimator,
\begin{equation}
    \hat s^\mathrm{id}_{QU} = \hat s^\mathrm{pd}_{QU} + (\tran{P_-} \Pi_{--} P_-)^{-1} \tran{P_-} \Pi_{-+} n_+ ,
\end{equation}
where $\Pi_{\pm\pm}$ is the $(\pm,\pm)$ block of the inverse transformed noise covariance matrix $(X N X)^{-1}$.
Whenever the noise properties of the two orthogonal detectors in a pair are identical, we have $N_{-+} = 0$ and thus $\Pi_{-+} = 0$ and $\hat s^\mathrm{id}_{QU} = \hat s^\mathrm{pd}_{QU}$.
As surprising as it may seem, this result merely emphasizes the fact that using all available data only improves on PD by modeling the noise cross-correlations between detectors of a same pair.
In this case, those are zero, such that the marginalization over the additional templates has no impact on the statistical precision of the map estimate.

We are now interested in quantifying the reduction of precision (excess variance) in PD relative to the ideal IQU estimator computed from all available data in more general cases, specifically when the noise properties of detectors in a pair can differ.
To examine this question, we now use atmosphere-free simulations, assuming that atmospheric contamination in the absence of (other) systematic effects has no impact on the recovered $Q$ and $U$ maps.
In these highly idealized conditions, our DW implementation is actually ideal since instrumental noise is simulated independently for each detector and is thus fully accounted for in the noise weights.
It thus provides a meaningful statistical benchmark to assess the performance of pair differencing.
We note that in more realistic conditions, the ``optimality'' of the DW approach as formulated in Sec.~\ref{sec:ml-solution} rests on several assumptions such as the Gaussian distribution of the atmospheric amplitudes $a$, which may not be a good model in general.
Pair differencing does not rely on such assumptions.

\subsection{Simulations}

We use simulations following almost the same setup as before, Sec.~\ref{sec:simulations}.
A small additional feature is the possibility to generate \emph{white} instrumental noise instead of $1/f$.
This is compatible with the detector-specific perturbations described previously.

\begin{figure}
    \includegraphics[width=\columnwidth]{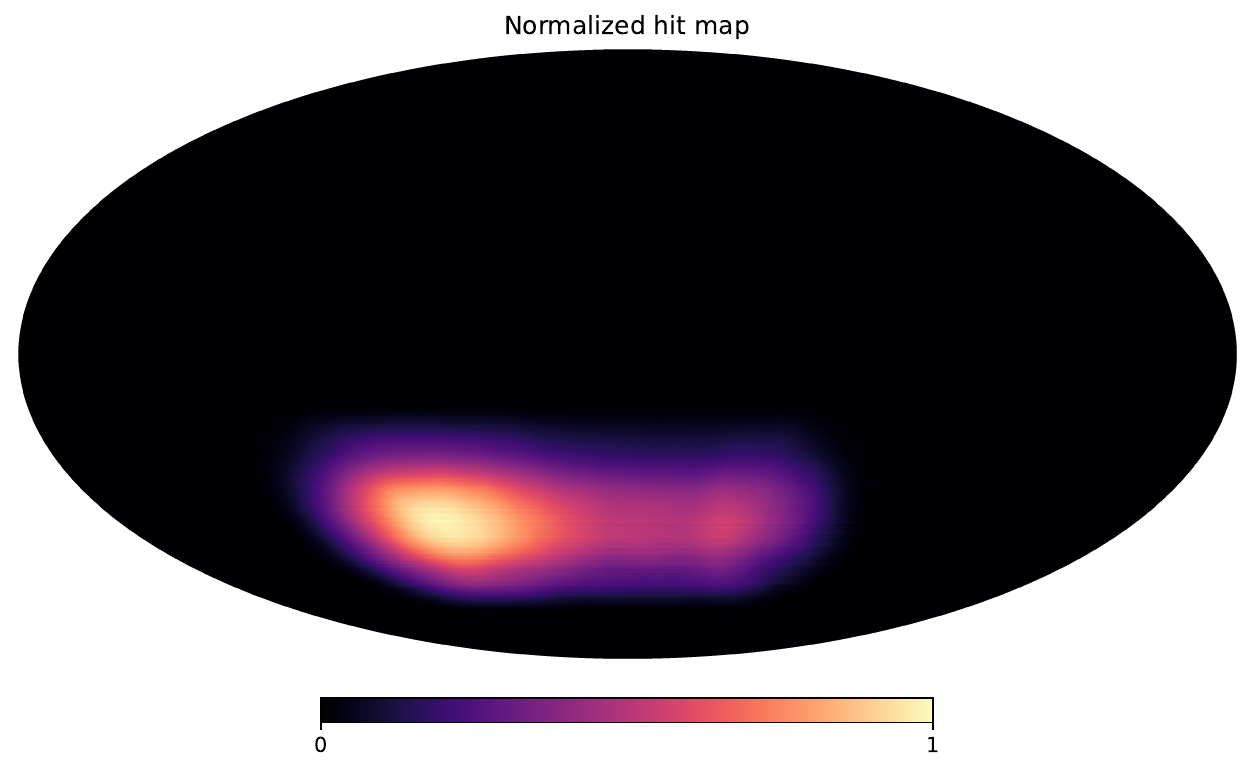}
    \caption{\label{fig:hitmap}
        Normalized hit map of the extended simulation (10 days of observation) in equatorial coordinates.
        The corresponding sky fraction is about $\qty{20}{\percent}$.
    }
\end{figure}

The other difference is a longer observing schedule covering not one, but 10 observing days, namely, the first day of each month except February and March, when weather conditions are typically degraded.
This schedule captures variations in the scanning pattern due to Sun and Moon avoidance across the year.
In total, those 10 days amount to 166 CESs, or approximately $\qty{137}{\hour}$ of observation.
The corresponding normalized hit map, showing the sky coverage of the simulation, is plotted in Fig.~\ref{fig:hitmap}.

\subsection{\label{sec:white-noise}Variance comparison at the map level in the white noise case}

The noise properties of the maps are contained in the pixel covariance matrix, $(\tran P N^{-1} P)^{-1}$.
This matrix is in general not accessible because of computational limitations.
However, when the instrumental noise is white, $N$ is diagonal, and the pixel covariance is simply the white noise covariance $\Delta$, Eq.~\eqref{eq:white-noise-covariance}.

In Appendix~\ref{sec:append-white-noise}, we compute analytically the \emph{excess variance} of the pair differencing estimate compared to the ideal estimate, considering a single detector pair, white instrumental noise, and noise levels independent between detectors.
Since PD does not take into account potentially different noise levels inside this single pair (but would optimally weight noise between different pairs), the excess variance in this case is simply a function of the relative difference of noise levels in the pair,
\begin{equation}
    \varepsilon \equiv \frac{\sigma_\parallel^2 - \sigma_\perp^2}{\sigma_\parallel^2 + \sigma_\perp^2} \in (-1, 1) ,
    \label{eq:epsilon}
\end{equation}
which is equal to zero when detectors have the same noise level and $\pm 1$ when one of them has zero (or infinite) noise.
The polarization covariance matrices in the pixel domain for this single pair are then related by
\begin{equation}
    \Delta^\text{id} = (1 - \varepsilon^2) \Delta^\text{pd} .
    \label{eq:var-increase-white}
\end{equation}
This confirms that the ideal estimate has lower variance than PD unless the noise levels are the same in both detectors.

In general, we define the relative excess variance as
\begin{equation}
    \zeta \equiv (\Delta^\text{pd} - \Delta^\text{id}) / \Delta^\text{id} .
    \label{eq:zeta}
\end{equation}
It is a random variable as it depends on the detector noise levels.
From Eq.~\eqref{eq:var-increase-white}, we see that $\zeta$ is the same in every map pixel for a single pair of detectors with given noise levels.
This will be our theoretical expectation for the excess variance:
\begin{equation}
    \zeta^\text{theory} \equiv \frac{\varepsilon^2}{1-\varepsilon^2} \geqslant 0 .
    \label{eq:zeta-theory}
\end{equation}
This may, however, not be true in the general case of multiple detector pairs because different focal plane pixels have different sky footprints (SATs typically have a field of view of several tens of degrees).
Additionally, the noise level of each detector can vary during the observing season.

Using the simulations described earlier, we have access to the actual $\zeta$ in each pixel of the map; therefore, we can study how the analytical result \eqref{eq:zeta-theory} generalizes to the case of multiple detector pairs and multiple independent CESs.
Here are the two main takeaways:
\begin{enumerate}[label=(\roman*)]
    \item The empirical average of $\zeta^\text{theory}$ over detector pairs and CESs is a good proxy for the mathematical expectation over all possible values of $\varepsilon$.
    \item The average $\expect{\zeta}_\text{pixels}$ across the map also follows this expectation.
\end{enumerate}

\begin{figure*}
    \includegraphics[width=0.7\textwidth]{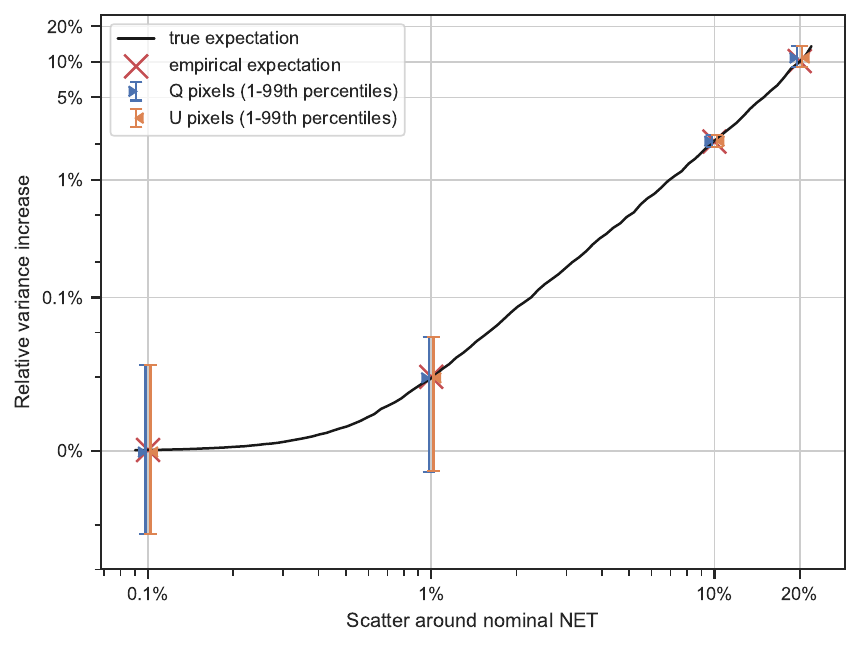}
    \caption{\label{fig:var-increase-white}
        Relative excess variance $\zeta$ in the white noise case as a function of the dispersion of noise levels $z$.
        The mathematical expectation (solid black) is computed numerically following Eqs.~\eqref{eq:zeta-theory} and \eqref{eq:epsilon} after drawing samples of $\sigma_\parallel^2$ and $\sigma_\perp^2$ from the same Gaussian distribution as the noise levels in the simulation.
        The empirical average (red crosses) over the detector pairs and CESs shows good agreement with the expectation at the four simulated values, $z \in \{\qty{0.1}{\percent}, \qty{1}{\percent}, \qty{10}{\percent}, \qty{20}{\percent}\}$.
        Error bars represent the 1st to 99th percentile of the $Q$ (blue) and $U$ (orange) pixel distribution of $\zeta$, with the triangle marks showing the average value, in excellent agreement between the two Stokes parameters.
        Blue and orange markers are slightly shifted left and right in order to improve visibility.
    }
\end{figure*}

The results are visualized in Fig.~\ref{fig:var-increase-white}, which shows the relative excess variance as a function of the dispersion of noise levels $z \in \{\qty{0.1}{\percent}, \qty{1}{\percent}, \qty{10}{\percent}, \qty{20}{\percent}\}$.
Unsurprisingly, we observe that $\zeta$ grows with $z$.
This illustrates how weighting individual detectors optimally in the ideal estimator takes advantage of the less noisy individual detectors to improve the map quality, while PD does not capitalize on this.
As $z$ increases, there are more and more pairs with a large discrepancy in noise levels, which intensifies the effect.

The agreement between empirical expectation and measured excess variance is useful from a practical point of view: it shows that we can estimate how the noise level of a map increases with respect to the ideal case using only the detector noise levels, which can be derived from the data directly, without going through the whole mapmaking procedure.
The 99th percentile of the pixel distribution of $\zeta$ provides a pessimistic estimate that would bound the excess variance for $\qty{99}{\percent}$ of the map area and at the same time account for deviations between the actual values and the empirical estimate from the detector noise levels.
Overall, the excess variance is \begin{enumerate*}[label=(\roman*)]
    \item under $\qty{0.1}{\percent}$ ($\qty{0.5}{\percent}$ pessimistic) when $z \lesssim \qty{1}{\percent}$,
    \item $\sim\qty{2}{\percent}$ ($\qty{2.5}{\percent}$ pessimistic) when $z = \qty{10}{\percent}$, and
    \item $\sim\qty{10}{\percent}$ ($\qty{15}{\percent}$ pessimistic) when $z = \qty{20}{\percent}$.
\end{enumerate*}

\subsection{Impact of variable instrumental noise parameters on the determination of the tensor-to-scalar ratio}

We now turn to the more realistic case of instrumental $1/f$ noise with variations from detector to detector as described in Sec.~\ref{sec:simulations}.
Contrary to the previous section, where we explored the white noise case, we do not have direct access to the exact map-space covariance matrix: it is a dense object because of temporal correlations in the noise and cannot in general be computed.
Therefore, we evaluate the loss of statistical power at the angular power spectrum level, using results from 25 different random noise realizations.

\begin{figure*}
    \includegraphics[width=\textwidth]{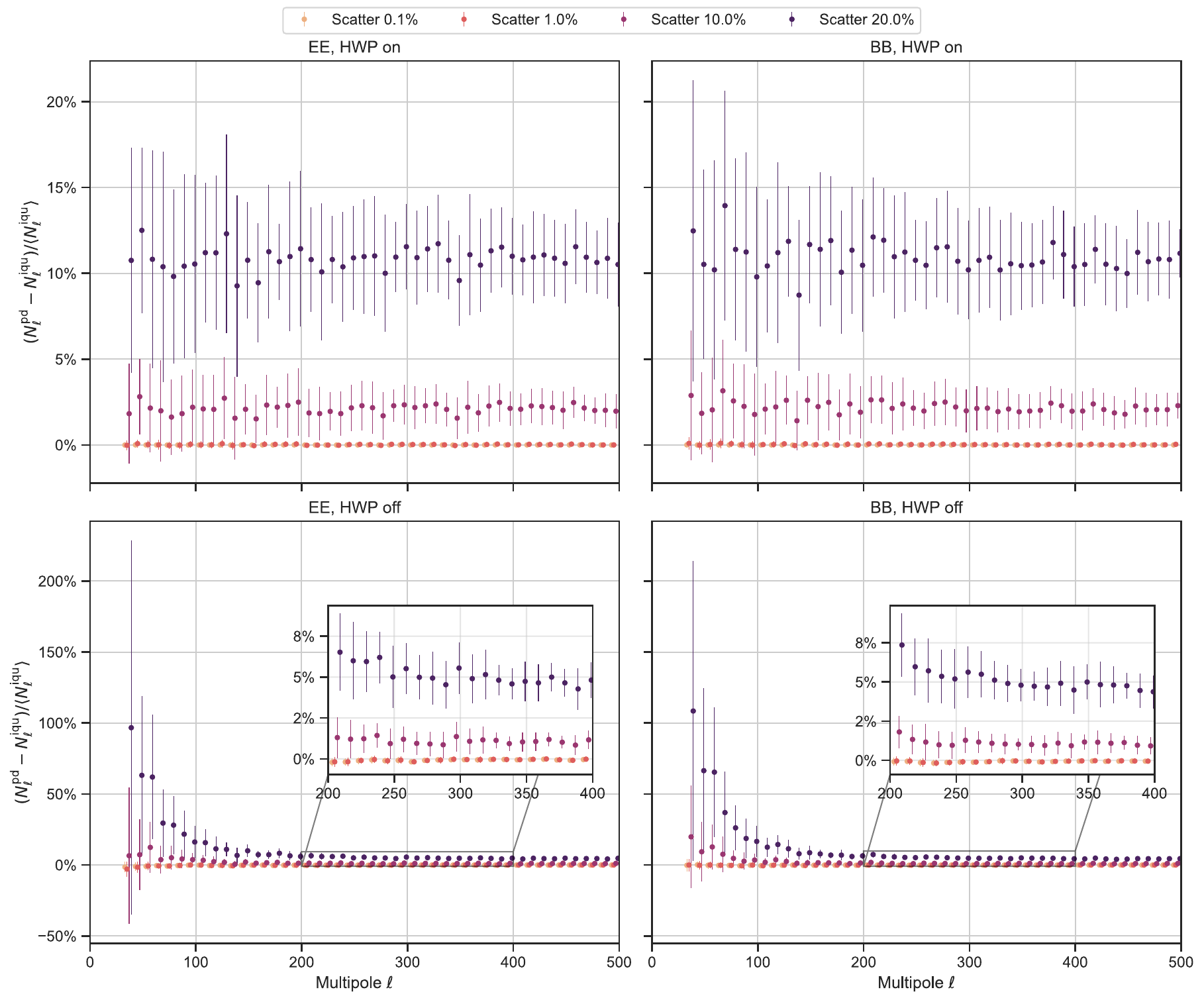}
    \caption{\label{fig:var-increase-instr}
        Increase of noise power in PD maps relative to the ideal high-$\ell$ white noise floor, computed as an average of the $\qty{10}{\percent}$ largest multipole bins.
        The dependence on the dispersion $z$ is encoded by color, and the markers and error bars, respectively, show the average and standard deviation over 25 independent instrumental noise realizations.
        Top panels have a rotating HWP, and bottom panels do not.
        Left (respectively, right) panels show the $EE$ (respectively, $BB$) spectra.
        A threshold of $\qty{10}{\percent}$ of the maximum hit count is applied to the maps, before apodizing them on a $\qty{10}{\degree}$ radius.
    }
\end{figure*}

\autoref{fig:var-increase-instr} shows the increase in PD noise spectra with respect to the ideal case, averaged over the 25 realizations, along with the standard deviation.
It illustrates the added value of a rotating HWP in mitigating large-scale correlated noise not rejected by the differencing operation.
While the two top panels show that the excess noise is ``white'' (does not depend on $\ell$) in the presence of a HWP, the noise curves in the bottom panels (no HWP) exhibit a $1 / \ell$-like behavior, with an important dependence on the variability of noise parameters in the detector pairs.

\begin{table*}
\begin{tabular}{dcccccc} \toprule
& \multicolumn{3}{c}{With HWP} & \multicolumn{3}{c}{Without HWP} \\
\cmidrule(rl){2-4}
\cmidrule(rl){5-7}
\multicolumn{1}{c}{Dispersion} & \multicolumn{1}{c}{$\sigma(r)^\text{id} [\times 10^{-3}]$} & \multicolumn{1}{c}{$\sigma(r)^\text{pd} [\times 10^{-3}]$} & \multicolumn{1}{c}{\textbf{Increase [\%]}} & \multicolumn{1}{c}{$\sigma(r)^\text{id} [\times 10^{-3}]$} & \multicolumn{1}{c}{$\sigma(r)^\text{pd} [\times 10^{-3}]$} & \multicolumn{1}{c}{\textbf{Increase [\%]}} \\
\midrule
 0.1 & $1.378\pm0.071$ & $1.378\pm0.071$ & $0.001\pm0.001$ & $3.786\pm0.489$ & $3.781\pm0.487$ & $-0.123\pm0.055$ \\
 1.0 & $1.378\pm0.071$ & $1.378\pm0.071$ & $0.017\pm0.008$ & $3.789\pm0.489$ & $3.784\pm0.487$ & $-0.111\pm0.048$ \\
10.0 & $1.363\pm0.067$ & $1.375\pm0.070$ & $0.889\pm0.152$ & $3.879\pm0.496$ & $3.947\pm0.507$ & $ 1.773\pm0.048$ \\
20.0 & $1.309\pm0.059$ & $1.361\pm0.068$ & $3.977\pm0.526$ & $4.075\pm0.507$ & $4.448\pm0.570$ & $ 9.161\pm0.418$ \\
\bottomrule
\end{tabular}
\caption{\label{tab:sigma-r-increase}
    Fisher uncertainty from 25 noise realizations on the tensor-to-scalar ratio $r$ for the ideal and PD estimators, assuming the true $r=0$, with and without a rotating HWP, for dispersion $z \in \{\qty{0.1}{\percent}, \qty{1}{\percent}, \qty{10}{\percent}, \qty{20}{\percent}\}$ of the noise parameters.
    In addition to the absolute numbers, the relative increase of uncertainty inherent to the PD estimator is given in percent.
}
\end{table*}

%

\autoref{tab:sigma-r-increase} summarizes those observations by quoting the Fisher uncertainty $\sigma(r=0)$ on the tensor-to-scalar ratio $r$, computed using
\begin{multline}
    \sigma(r=0)^{-2} \approx \frac{f_\text{sky}}{2} \\
    \times \sum_{\text{bins $b$}} \delta_\ell (2\ell_b+1) \Bigg( \frac{C_{\ell_b}^{BB,\text{prim}} \rvert_{r=1}}{C_{\ell_b}^{BB,\text{lens}} + \expect{N_{\ell_b}^{BB}}} \Bigg)^2
    \label{eq:fisher}
\end{multline}
where $\expect{N_{\ell_b}^{BB}}$ is the average noise power spectrum over the 25 noise realizations, and $\delta_\ell = 10$ stands for the bin size.
The table also gives for each value a $1\sigma$ confidence interval computed by evaluating Eq.~\eqref{eq:fisher} at the average value of the noise spectrum, plus/minus the standard deviation over the 25 noise realizations.
As expected, the increase of uncertainty is more pronounced when no HWP is used, as the power spectrum estimates at low multipoles are more noisy, and this is where most of the signal of interest is.
For typical dispersion values $z \sim \qty{10}{\percent}-\qty{20}{\percent}$, the uncertainty on $r$ could increase by a few percent when using a HWP, and as much as $\qty{10}{\percent}$ when not.

One may notice two potentially surprising features in \autoref{tab:sigma-r-increase}.
First, the absolute uncertainty on $r$ in the case with HWP actually \emph{decreases} with higher values of $z$.
This is because, given the implementation of the perturbed noise model, having larger dispersion of the noise parameters around their nominal values does not necessarily mean that the detector array loses sensitivity overall.
In the case without HWP, this effect is buried under the excess noise at low multipoles.

The second feature is that the PD estimator seems to perform better than the ideal estimator, in the no HWP case, for low $z$ values.
This seems inconsistent with the fact that, from first principles, the ideal estimate should have minimal variance.
It turns out that this is driven by the very low-multipole bins ($\ell < 50$).
However, inspecting the scatter of their values across the 25 realizations reveals that each bin is statistically consistent with the ideal case.
We therefore consider that this is a statistical fluke and/or caused by uncontrollable numerical effects that could be due too many different factors, such as imperfections in the estimation of the noise weights, estimation of the pseudospectra, etc.
Finally, we emphasize that this is anyway a very small effect (about $\qty{0.1}{\percent}$) and does not affect our conclusions.

\section{\label{sec:systematics} Pair differencing in the presence of instrumental systematic effects}

In previous sections, we demonstrated two key points about pair differencing.
First, it provides estimates of the polarized sky that are as optimal as possible when arbitrary, unpolarized common modes are present.
Second, in practical applications, PD maps nearly have the best possible precision given the sensitivity of the instrument itself, i.e., as if atmospheric correlated noise were not present.
While these results are promising, they assumed rather idealized conditions, without instrumental systematics.
All mapmaking methods are affected by those, though some handle certain types of systematics better than others.
The specific impact may depend heavily on the particular systematic effect and instrument involved, requiring case-by-case analysis.
The purpose of this section is not to replace such detailed studies, but rather to show that PD could still perform well, and deserves consideration in real-world analysis pipelines.

We consider here the simple example of \emph{gain mismatch} between the two detectors of a pair \cite{Shimon:2007au,POLARBEAR:2014hgp,BICEP2:2014owc}.
This undermines the ability of PD to cleanly remove signals common to both detectors, a key strength of the method.
Still, given that relative calibration factors in real-world experiments can typically be determined with percent-level accuracy \cite{ACT:2025xdm}, we expect most of the atmospheric contamination to be rejected by the differencing operation.
Any residual leakage is then dealt with by using appropriate weights reflecting the additional $1/f$ noise.

To investigate the impact of gain mismatch on the recovered PD maps, we use the setup described in Sec.~\ref{sec:simulations}.
For each detector pair, we multiply the atmosphere contribution to the data of the first detector by $1+\delta/2$ and that of the second one by $1-\delta/2$, where $\delta$ is drawn randomly around zero with some dispersion (either 0.1 or $\qty{1}{\percent}$) and independently for each detector pair.
After forming the difference streams, a $1/f$ model, Eq.~\eqref{eq:noise-model}, is fitted to each of them in order to account for both the instrumental noise and residual atmosphere.
The maps are then estimated as in Eq.~\eqref{eq:pd-estimate-gen}.

\begin{figure*}
    \includegraphics[width=\textwidth]{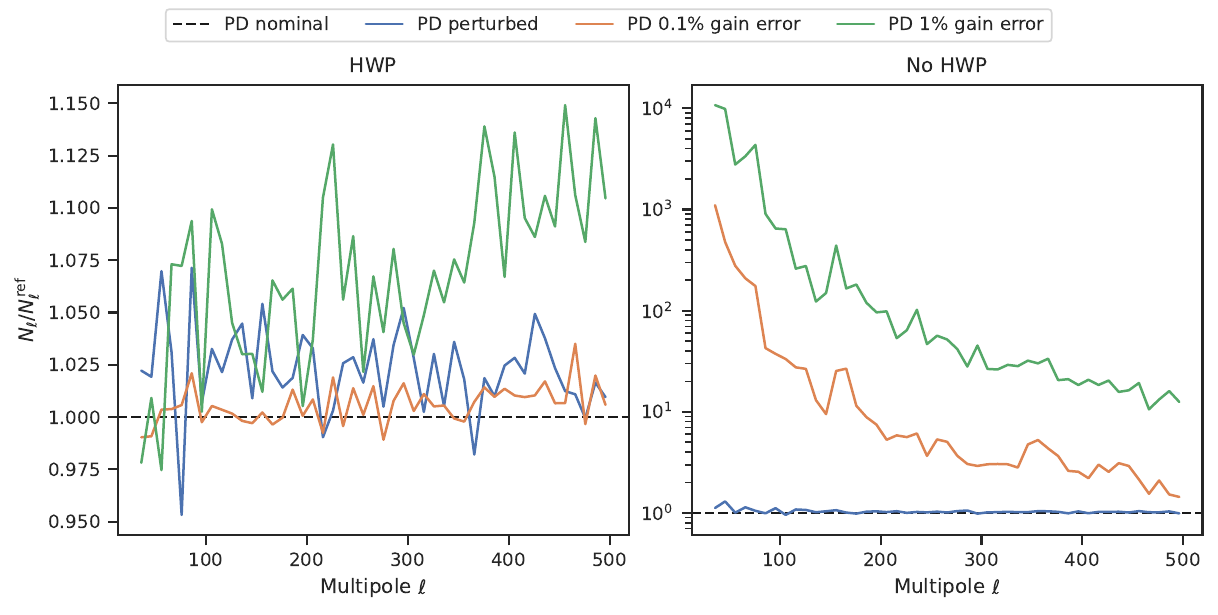}
    \caption{\label{fig:leakage}
        Comparison of $BB$ noise spectra using pair differencing, with (left) and without (right) HWP rotation.
        Spectra are plotted relative to a reference case with no systematics and instrumental noise parameters at there nominal values.
        The ``perturbed'' curves (blue) correspond to a typical $\qty{10}{\percent}$ variation around the nominal values.
        Finally, cases with $\qty{0.1}{\percent}$ (respectively, $\qty{1}{\percent}$) gain errors are plotted in orange (respectively, green).
    }
\end{figure*}

In Fig.~\ref{fig:leakage}, left panel, we show that HWP modulation still enables a precise recovery of the polarization information.
Indeed, the additional variance at large angular scales caused by the leakage seems of comparable scale to the one due to a variation of instrumental noise parameters across the focal plane (cf. Sec.~\ref{sec:sensitivity}).
When no HWP is used (right panel), the situation is different.
Atmospheric leakage increases the variance by roughly 2 to 4 orders of magnitude at low multipoles ($\ell < 100$) depending on the level of gain mismatch.
It is likely that better modeling of the residual atmospheric noise, in particular inclusion of the correlations between detector pairs, could lead to more precise recovery of the polarization signal in both cases.
Nonetheless, we conclude overall that even in the presence of such residuals, owing to the presence of a continuously rotating HWP, the simple PD technique, as is, is capable of delivering maps of quality comparable to the best achievable while making minimal assumptions about atmospheric contamination.

More complex sources of systematic effects will unavoidably affect the data of real experiments and impact PD performance, for instance, slightly different or asymmetric beams \cite{Shimon:2007au}.
We expect that the powerful combination of orthogonal detector pairs and HWP will continue helping to mitigate many of them, as long as their main consequence is the atmospheric, or more generally any unpolarized, signal leakage to the difference data.
Other classes of effects may instead require extensions of the formalism, for example, the introduction of additional degrees of freedom to model them, e.g., Refs.~\cite{POLARBEAR:2014hgp, McCallum:2020jsp}, thus potentially also affecting the overall precision of the recovered maps.
This needs, however, to be studied in detail, case by case, for PD as well as any other mapmaking procedure. 

In this work, we assumed an \emph{ideal} HWP.
But, in reality, no optical component is perfect, and HWP nonidealities could diminish its ability to mitigate the unpolarized low-frequency noise \cite{Monelli:2023wmv, Patanchon:2023ptm}.
In particular, intensity-to-polarization effects can modulate a fraction of the atmospheric signal to frequencies that correspond to harmonics of the HWP rotation frequency.
The impact of such HWP-related systematics on the PD method and their interaction with other instrumental effects is left for future work.

\section{\label{sec:conclusion}Conclusion}

In this work, we have compared two mapmaking approaches based on their ability to precisely reconstruct CMB polarization maps from ground-based experiments in the presence of atmospheric contamination.
In particular, we have demonstrated that pair differencing, a technique that relies on the rejection of common-mode noise using pairs of orthogonally polarized detectors, emerges naturally as a maximum-likelihood estimator under the assumption of perfectly correlated atmospheric emission between detectors in a pair, without explicit modeling of the contaminant signal.
This is a key feature given the dynamic and turbulent nature of atmospheric fluctuations, which makes them difficult to characterize and model accurately.

Our results show that PD achieves close to ideal sensitivity to the inflationary tensor-to-scalar ratio $r$ in the presence of realistic variations of instrumental noise properties across the focal plane, despite using only half of the available data and thus neglecting cross-correlations between detectors of a pair.
In particular, we find only a small (few percent) enlargement of the uncertainty on $r$ compared to an idealized reconstruction from atmosphere-free data, when a rotating half-wave plate is used.
This degradation increases to about $\qty{10}{\percent}$ in the absence of a HWP.

Compared to usual downweighting approaches, which include both instrumental and atmospheric noise in the noise weights, PD offers a simpler and more numerically stable alternative, particularly when detectors in a pair have similar noise properties.
Moreover, in our tests, we find that DW typically has to perform an \emph{implicit} differencing operation, by assigning the same weights to both detectors in a pair, to reach competitive performance with PD.
Since the atmosphere is often used as a beam-filling calibrator to compute relative gain factors, we expect that this remains the case in practice, unless a very accurate model of the atmospheric noise is available, allowing for its removal from the data.
We also note that the inclusion of cross-correlations between different detectors in the noise model should allow DW to better mitigate atmospheric contamination while taking into account variations within detector pairs.

We conclude that for the recovery of polarization maps, and experiments featuring both a HWP and pairs of orthogonal detectors, PD is a numerically efficient method that is surprisingly good at mitigating unpolarized atmospheric contamination in a model-independent way, while delivering nearly optimal performance in terms of statistical uncertainty of the recovered maps and potentially being as vulnerable as any other method to some common systematic effects.
Future work should presumably explore hybrid strategies and systematic mitigation schemes to further enhance the robustness and accuracy of polarization reconstruction.

\begin{acknowledgments}
The authors would like to thank Hamza El Bouhargani, Wuhyun Sohn and the SciPol team for useful discussions.

S.B. acknowledges partial Ph.D. funding from the DIM ORIGINES program (project RADIATION).
This work was supported by the SCIPOL project,\footnote{\url{https://scipol.in2p3.fr}} funded by the European Research Council (ERC) under the European Union’s Horizon 2020 research and innovation program (PI: Josquin Errard, Grant No.~101044073).
The authors also benefited from the European Union’s Horizon 2020 research and innovation program under Grant No.~101007633 CMB-Inflate.

This work was granted access to the HPC resources of IDRIS under the allocation 2024-AD010414161R2 and 2025-AD010416919R1 made by GENCI.
This research also used resources of the National Energy Research Scientific Computing Center (NERSC), a Department of Energy User Facility using NERSC award HEP-ERCAP 0034243.
Results presented in this paper have made use of the following packages: \texttt{NaMaster} \cite{Alonso:2018jzx}, \texttt{healpy} \cite{Zonca:2019vzt}, \texttt{NumPy} \cite{Harris:2020xlr}, \texttt{SciPy} \cite{Virtanen:2019joe}, \texttt{Matplotlib} \cite{Hunter:2007ouj}, and \texttt{seaborn} \cite{Waskom:2021psk}.

\end{acknowledgments}

\appendix
\section{\label{sec:append-marginalization}Pair differencing as template marginalization}

Assume a data model where two orthogonal detectors, denoted by $\parallel$ and $\perp$, measure the same, unknown, total intensity signal $a$ in addition to sky polarization $s_{QU}$:
\begin{equation}
    d = \begin{bmatrix} d_\parallel \\ d_\perp \end{bmatrix}
    = \begin{bmatrix} T \\ T \end{bmatrix} a
    + \begin{bmatrix} P_{QU} \\ -P_{QU} \end{bmatrix} s_{QU}
    + n .
\end{equation}
Using the pair transformation $
    X = \frac12 \begin{bsmallmatrix*}[r]
        \mathbb I & \mathbb I \\
        \mathbb I & -\mathbb I
    \end{bsmallmatrix*}
$ from Eq.\eqref{eq:transformation-matrix}, the data model is rewritten as
\begin{equation}
    X d = \begin{bmatrix} d_+ \\ d_- \end{bmatrix}
    = \underbrace{\begin{bmatrix} T \\ 0 \end{bmatrix}}_{\equiv \mathcal T} a
    + \underbrace{\begin{bmatrix} 0 \\ P_{QU} \end{bmatrix}}_{\equiv \mathcal P} s_{QU}
    + X n .
    \label{eq:transformed-data-model}
\end{equation}

The sky polarization estimate that deprojects the total intensity signal (and corresponds to marginalizing over the total intensity amplitudes $a$) is
\begin{subequations}
\begin{align}
    \hat s_{QU} &= (\tran{\mathcal P} F \mathcal P)^{-1} \tran{\mathcal P} F (Xd) \\
\intertext{with}
    F &\equiv C - C \mathcal T (\tran{\mathcal T} C \mathcal T)^{-1} \tran{\mathcal T} C \label{eq:filter} \\
    C &\equiv (X N X)^{-1} .
\end{align}
\end{subequations}

We want to minimize the assumptions on the total intensity signal.
This amounts to deprojecting anything that is measured the same way by both detectors, i.e., taking the original $T = \mathbb I$.
In particular, we do not assume that this contribution can be pixelized or has any predictable structure.
We now show that this leads to the pair differencing estimator.

Given the structure of $\mathcal P = P_{QU} \begin{bsmallmatrix}
    0 \\ \mathbb I
\end{bsmallmatrix}$ and $\mathcal T = \begin{bsmallmatrix}
    \mathbb I \\ 0
\end{bsmallmatrix}$, defined in Eq.~\eqref{eq:transformed-data-model}, any sandwiches between those matrices extract one of the four blocks of the object in the middle.
For example, the template orthogonalization kernel is simply
\begin{equation}
    \tran{\mathcal T} C \mathcal T = \begin{bmatrix}
        \mathbb I & 0
    \end{bmatrix} \begin{bmatrix}
        C_{11} & C_{12} \\ C_{21} & C_{22}
    \end{bmatrix} \begin{bmatrix}
        \mathbb I \\ 0
    \end{bmatrix}
    = C_{11} .
\end{equation}
The mapmaking kernel, on the other hand, is
\begin{equation}
    \tran{\mathcal P} F \mathcal P = \underbrace{\begin{bmatrix}
        \mathbb I & 0
    \end{bmatrix} \tran{P_{QU}}}_\text{commute} \begin{bmatrix}
        F_{11} & F_{12} \\ F_{21} & F_{22}
    \end{bmatrix} \underbrace{P_{QU} \begin{bmatrix}
        \mathbb I \\ 0
    \end{bmatrix}}_\text{commute}
    = \tran{P_{QU}} F_{22} P_{QU} .
\end{equation}
Using this trick, we can write the estimator as
\begin{equation}
    \hat s_{QU} = (\tran{P_{QU}} F_{22} P_{QU})^{-1} \tran{P_{QU}} (F_{21} d_+ + F_{22} d_-) .
\end{equation}
By definition of the filtering operator, we have $F \mathcal T = 0$.
Thus, $F_{11} = F_{21} = 0$.
All that is left is to compute $F_{22}$.
This is straightforward from its definition \eqref{eq:filter},
\begin{equation}
\begin{split}
    F_{22} &= \left(C - C_{\cdot 1} (C_{11})^{-1} C_{1 \cdot}\right)_{22} \\
    &= C_{22} - C_{2 1} (C_{11})^{-1} C_{1 2} \\
    &= \left((C^{-1})_{22}\right)^{-1} \quad \text{by blockwise inversion} \\
    &= \left((XNX)_{22}\right)^{-1} ,
\end{split}
\end{equation}
and we recognize the inverse noise covariance matrix of the difference data.
Therefore, the estimator is simply the pair differencing estimator \eqref{eq:pd-estimate}, which concludes the proof.

\section{\label{sec:append-iqu-pol-estimator}Polarized part of the full IQU estimator}

\textcite[Appendix C]{ElBouhargani:2021phd} shows that, when the instrumental noise is the same in two detectors of a pair, the pair differencing solution is \emph{as optimal} as the simultaneous $IQU$ reconstruction.
We broadly recall the argument here but refer anyone interested in the detailed derivation to the original work.

Let us introduce the pair-sum and pair-difference data vectors and associated pointing matrices and noise vectors,
\begin{subequations}
\begin{align}
    d_+ &= \frac12 \left( d_\parallel + d_\perp \right)
    = P_+ s + n_+\\
    d_- &= \frac12 \left( d_\parallel - d_\perp \right)
    = P_- s + n_-
    \label{eq:diffStream}
\end{align}
\end{subequations}
where the $\parallel$ and $\perp$ subscripts denote the orthogonal detectors of a pair.

The $IQU$ solution $\hat s$ can be rewritten in terms of the transformed dataset, with block versions of the matrices,
\begin{equation}
\begin{split}
    \hat s
    &= \left( \tran{P} N^{-1} P \right)^{-1} \tran{P} N^{-1} d \\
    &= \left( \begin{bmatrix} \tran{P_+} & \tran{P_-} \end{bmatrix} \mathcal N^{-1} \begin{bmatrix} P_+ \\ P_- \end{bmatrix} \right)^{-1} \begin{bmatrix} \tran{P_+} & \tran{P_-} \end{bmatrix} \mathcal N^{-1} \begin{bmatrix} d_+ \\ d_- \end{bmatrix} ,
\end{split}
\end{equation}
with the transformed noise covariance matrix $\mathcal N$ inverted blockwise:
\begin{equation}
    \mathcal N^{-1}
    \equiv \begin{bmatrix} N_{++} & N_{+-} \\ N_{-+} & N_{--} \end{bmatrix}^{-1}
    \equiv \begin{bmatrix} \Pi_{++} & \Pi_{+-} \\ \Pi_{-+} & \Pi_{--} \end{bmatrix}
    .
    \label{eq:transformed-noise-covariance}
\end{equation}
$\Pi_{\square\square}$ is just a notation for the $\square\square$ block of $\mathcal N^{-1}$.

From there, we can write the map estimator in block form, separating the intensity and polarization components.
The polarized part is given by Ref.~\cite{ElBouhargani:2021phd}, Eq.~(C.14),
\begin{equation}
    \hat s_{QU} = s_{QU} + \left( \tran{P_-} F_{--} P_- \right)^{-1} \tran{P_-} \left[ F_{--} n_- + F_{-+} n_+ \right]
    \label{eq:iqu-pol-estimator}
\end{equation}
with
\begin{subequations}
    \begin{align}
        F_{--} &= \Pi_{--} - \Pi_{-+} P_+ (\tran{P_+} \Pi_{++} P_+)^{-1} \tran{P_+} \Pi_{+-} \\
        F_{-+} &= \Pi_{-+} - \Pi_{-+} P_+ (\tran{P_+} \Pi_{++} P_+)^{-1} \tran{P_+} \Pi_{++} .
    \end{align}
\end{subequations}

We see from Eq.~\eqref{eq:iqu-pol-estimator} that $\hat s_{QU}$ is the sum of three terms:
\begin{enumerate*}[label=(\roman*)]
    \item the true sky map,
    \item a direct contribution from the pair-difference noise, and 
    \item a cross-contribution from the pair-sum noise.
\end{enumerate*}
Whenever the covariance of the pair-sum and pair-difference noise vectors, $N_{+-}$, is small, the transformed noise covariance matrix $\mathcal N$ becomes block-diagonal,
\begin{equation}
    \mathcal N = \begin{bmatrix}
        N_{++} & \simeq 0 \\
        \simeq 0 & N_{--}
    \end{bmatrix},
\end{equation}
such that
\begin{equation}
    F_{-+} = 0 \quad \text{and} \quad F_{--} = N_{--}^{-1} .
\end{equation}
In this limit, the optimal polarization map reduces to the PD solution \eqref{eq:pd-estimate}:
\begin{equation}
    \hat s_{QU} \xrightarrow{N_{-+} \to 0} \hat s^\text{pd}_{QU} = s_{QU} + (\tran{P_-} N_{--}^{-1} P_-)^{-1} \tran{P_-} N_{--}^{-1} n_- .
\end{equation}

\section{\label{sec:append-white-noise}Pair differencing in the case of white noise}

We start by simplifying Eq.~\eqref{eq:iqu-pol-estimator} by assuming that all sky pixels are observed with a uniform variety of crossing angles, such that
\begin{subequations}
    \begin{align}
        \tran P_- F_{--} &\approx \tran P_- \Pi_{--} \\
        \tran P_- F_{-+} &\approx \tran P_- \Pi_{-+}
    \end{align}
    \label{eq:uniform-angle-sampling}
\end{subequations}
This simplification is a very good approximation for the deepest parts of the map, which are observed many times by different detectors with varying telescope orientations.
The presence of a rotating HWP makes this case compelling.
The $QU$ part of the estimate now reads
\begin{equation}
    \hat s_{QU} = s_{QU} + \left( \tran{P_-} \Pi_{--} P_- \right)^{-1} \tran{P_-} \left[ \Pi_{--} n_- + \Pi_{-+} n_+ \right] .
    \label{eq:estimator-uniform-coverage}
\end{equation}

We denote the white noise levels of the even- and odd-index detectors by $\sigma_\parallel$ and $\sigma_\perp$, respectively.
They are free to vary from one detector pair to another, and we treat them as independent Gaussian random variables.
Leaving out any correlations between different detectors, we consider a \emph{single pair} of detectors with independent noise levels following the same distribution $\mathcal N(\bar\sigma, z^2)$ centered on a nominal value $\bar\sigma$ and with variance $z^2$.
We will comment on the multidetector case at the end.

These assumptions simplify the starting equation \eqref{eq:estimator-uniform-coverage} because the noise matrices are just scalars ($\propto \mathbb I$), so they commute with other matrices:
\begin{equation}
\begin{split}
    \hat s_{QU}
    &= s_{QU} + \underbrace{\left(\tran{P_-} P_-\right)^{-1} \tran{P_-}}_{\equiv B_-}
    \Bigl[n_- + \underbrace{(\Pi_{--})^{-1} \Pi_{-+} n_+}_{\equiv \tilde n_+}\Bigr] \\
    &= \underbrace{s_{QU} + B_- n_-}_{\text{PD estimator}} + B_- \tilde n_+ .
\end{split}
\label{eq:white-noise-estimator}
\end{equation}
We have recognized the PD estimator $\hat s_{QU}^\text{pd}$ and labeled the binning operator $B_-$.

The blocks of the transformed noise covariance matrix $\mathcal N$ \eqref{eq:transformed-noise-covariance} are related to those of the original one by
\begin{subequations}
\begin{align}
    N_{++} &= \frac14 \left( N_\parallel + 2 N_{\parallel\times\perp} + N_\perp \right) = \frac14 \left(\sigma_\parallel^2 + \sigma_\perp^2\right) \mathbb I \\
    N_{--} &= \frac14 \left( N_\parallel - 2 N_{\parallel\times\perp} + N_\perp \right) = \frac14 \left(\sigma_\parallel^2 + \sigma_\perp^2\right) \mathbb I \\
    N_{+-} &= N_{-+} = \frac14 \left( N_\parallel - N_\perp \right) = \frac14 \left(\sigma_\parallel^2 - \sigma_\perp^2\right) \mathbb I
\end{align}
\end{subequations}
with $N_{\parallel\times\perp} = 0$ in our case where detector noise levels are uncorrelated.
One can write the inverse noise covariance blocks $\Pi$ appearing in Eq.~\eqref{eq:white-noise-estimator} explicitly, omitting the identity matrix,
\begin{align*}
    (\Pi_{--})^{-1}
    &= N_{++} - N_{+-} N_{--}^{-1} N_{-+} \\
    &= \begin{multlined}[t]
        \frac14(\sigma^2_\parallel + \sigma^2_\perp) \\
        - \frac14(\sigma^2_\parallel - \sigma^2_\perp) \times
        \frac{4}{\sigma^2_\parallel + \sigma^2_\perp} \times
        \frac14(\sigma^2_\parallel - \sigma^2_\perp)
    \end{multlined} \\
    &= \frac{\sigma^2_\parallel \sigma^2_\perp}{\sigma^2_\parallel + \sigma^2_\perp} ,
\intertext{and similarly,}
\begin{split}
    \Pi_{-+}
    &= - \Pi_{--} N_{-+} N_{++}^{-1} \\
    &= - \frac{\sigma^2_\parallel + \sigma^2_\perp}{\sigma^2_\parallel \sigma^2_\perp}
    \times \frac14(\sigma^2_\parallel - \sigma^2_\perp)
    \times \frac{4}{\sigma^2_\parallel + \sigma^2_\perp} \\
    &= - \frac{\sigma^2_\parallel - \sigma^2_\perp}{\sigma^2_\parallel \sigma^2_\perp} .
\end{split}
\end{align*}
Together, these give the cross-noise term
\begin{equation}
    \tilde n_+ = -\frac{\sigma^2_\parallel - \sigma^2_\perp}{\sigma^2_\parallel + \sigma^2_\perp} n_+ .
\end{equation}

It is now straightforward to compute the pixel-domain covariance matrix for each noise term and their cross-covariance.
We use $\Delta$ to denote those quantities, consistently with the main text, Eq.~\eqref{eq:white-noise-covariance}, and obtain
\begin{subequations}
\begin{align}
    \Delta_1 &\equiv \Var[B_- n_-]
    = \frac{\sigma^2_\parallel + \sigma^2_\perp}{4} \Sigma_-^{-1} \\
    \Delta_2 &\equiv \Var[B_- \tilde n_+]
    = \left(
    \frac{\sigma^2_\parallel - \sigma^2_\perp}{\sigma^2_\parallel + \sigma^2_\perp}
    \right)^2 \Delta_1
    = \varepsilon^2 \Delta_1 \\
    \Delta_{12} &\equiv \Cov[B_- n_-, B_- \tilde n_+]
    = - \Delta_2 ,
\end{align}
\label{eq:delta-matrices}
\end{subequations}
where
\begin{equation}
    \Sigma_- \equiv \tran{B_-} B_- = \tran{P_-} P_-
\end{equation}
describes the sky coverage in $Q$ and $U$ and $\varepsilon$ is the relative difference of noise levels in the detector pair, defined in Eq.~\eqref{eq:epsilon}, which is equal to zero when detectors have the same noise level and $\pm 1$ when one of them has zero (or infinite) noise.
Since $\Sigma_-$ is positive definite, $\Delta_1$ is as well, whereas the cross-covariance $\Delta_{12}$ is negative definite.
Therefore, at least in our simplified white noise model, the two noise terms of the $IQU$ estimator are \emph{anticorrelated} in every pixel.
This is important because we know that this estimator should be optimal (in the sense of minimum variance) and therefore have a smaller variance than the PD estimator which only has the first term with variance $\Delta_1$.
Quantitatively, we have
\begin{equation}
\begin{split}
    \Delta^\text{iqu} \equiv \Var[\hat s_{QU}]
    &= \Var[B_- n_- + B_- \tilde n_+] \\
    &= \Delta_1 + \Delta_2 + 2 \Delta_{12} \\
    &= \Delta_1 - \Delta_2 \\
    &= (1 - \varepsilon^2) \Delta_1 \\
    &\leqslant \Delta_1 = \Var[\hat s_{QU}^\text{pd}] \equiv \Delta^\text{pd} .
\end{split}
\label{eq:pixel-noise-covariance}
\end{equation}
The variance of the PD estimator is indeed larger than what can be obtained by computing the full IQU solution in the white noise case.
The increase of variance is given by the factor, uniform across the map,
\begin{equation}
    \eta \equiv \frac{\Delta^\text{pd}}{\Delta^\text{iqu}} = \frac{\Delta_1}{\Delta_1 - \Delta_2} = \frac{1}{1 - \varepsilon^2} \geqslant 1 ,
    \label{eq:increase-variance}
\end{equation}
with $\eta = 1$ when the noise levels of the two detectors are equal ($\varepsilon = 0$).

We emphasize that $\eta$ is a random variable, because it ultimately depends on the detector noise levels which are drawn randomly.
The expected increase of variance can be evaluated numerically by drawing many pairs $(\sigma_\parallel, \sigma_\perp)$ and averaging the corresponding $\eta$ values.

Let us now comment on the multidetector case.
The generalization of the result \eqref{eq:increase-variance} is not straightforward.
Each focal plane pixel observes a slightly different part of the sky, so the $\Sigma_-$ matrix is specific to each detector pair.
For example, the total covariance matrix for the PD estimator would be
\begin{equation}
    \Delta^\text{pd tot} = \left(\sum_\text{pairs} w \Sigma_-\right)^{-1}
    \; \text{with} \; w = \frac{4}{\sigma^2_\parallel + \sigma^2_\perp} .
\end{equation}
Because of the dispersion of noise levels across the focal plane, combined with slightly different footprints of those detectors (especially when considering an instrument with a wide field of view), this expression can not be simplified in the general case.

\bibliography{paper}

\end{document}